%% file: main.tex
\documentclass[lettersize,journal]{IEEEtran}
\usepackage{amsmath,amsfonts}
\usepackage{algorithmic}
\usepackage{algorithm}
\usepackage{booktabs} 
\usepackage{array}
\usepackage{longtable}
\usepackage{supertabular}
\usepackage{subcaption}
\usepackage{tabularx}
\usepackage{float}
\usepackage{graphicx}
\usepackage[referable]{threeparttablex}
\usepackage{textcomp}
\usepackage{stfloats}
\usepackage{url}
\usepackage{verbatim}
\usepackage{pifont}
\usepackage{cite}
\usepackage{svg}
\usepackage{dirtytalk}
\usepackage{hyperref}
\usepackage{multirow}
\usepackage{graphicx}

\newcolumntype{H}[1]{>{\hsize=#1\hsize\arraybackslash}X} 
\usepackage[resetlabels,labeled]{multibib}
\newcites{P}{Selected Primary Studies (PSs)}

\usepackage[inline]{enumitem}
\usepackage[many]{tcolorbox}
\newtcolorbox{boxD}{
    colback = white, 
    colframe = black, 
    boxrule = 0pt, 
    toprule = 1pt, 
    bottomrule = 1pt 
}
\definecolor{EF7F5A}{HTML}{EF7F5A}
\definecolor{FFD166}{HTML}{FFD166}
\definecolor{4EA654}{HTML}{4EA654}
\definecolor{1FB7EA}{HTML}{1FB7EA}
\definecolor{052B38}{HTML}{052B38}

\usepackage{xcolor} 
\newcommand{\revised}[1]{\textcolor{black}{#1}}
\usepackage[normalem]{ulem}

\makeatletter
\usepackage{amssymb}
\newcommand{\ygg@basicalert}[2]{{\bfseries\sffamily\scriptsize#1}{\sf\small$\blacktriangleright$\textit{#2}$\blacktriangleleft$}}
\newcommand{\YANN}[1]{\ygg@basicalert{YANN}{#1}}
\newcommand{\rqone}{ Which migration phases are automated by ML?}
\newcommand{\rqtwo}{ How are ML migration inputs characterised in terms of type, granularity, source, and preprocessing?}
\newcommand{\rqthree}{ What ML techniques and learning paradigms are applied in migration approaches?}
\newcommand{\rqfour}{ How are ML-based migration approaches evaluated with respect to metrics, benchmarks, tools, and success criteria?}
\newcommand{\rqfive}{ What challenges arise in using ML approaches?}
\begin{document}

\title{A Systematic Literature Review of Machine Learning Approaches for Migrating  Monolithic Systems to Microservices}

\author{Imen Trabelsi, Brahim Mahmoudi, Jean Baptiste Minani, Naouel Moha, and Yann-Ga\"{e}l Gu\'eh\'eneuc
\IEEEcompsocitemizethanks{
\IEEEcompsocthanksitem Imen Trabelsi is with the Department of Computer Science, École de Technologie Supérieure (ÉTS) – Université du Québec, Canada\protect\\
E-mail: iman.trabelssi@gmail.com
\IEEEcompsocthanksitem Brahim Mahmoudi is with the Department of Computer Science, École de Technologie Supérieure (ÉTS) – Université du Québec, Canada\protect\\
E-mail: brahim.mahmoudi.1@ens.etsmtl.ca
\IEEEcompsocthanksitem Jean Baptiste Minani is with the Department of Computer Science and Software Engineering, Concordia University, Canada\protect\\
E-mail: baptiste2k8@gmail.com
\IEEEcompsocthanksitem Naouel Moha is with the Department of Computer Science, École de Technologie Supérieure (ÉTS) – Université du Québec, Canada\protect\\
E-mail: naouel.moha@etsmtl.ca
\IEEEcompsocthanksitem Yann-Ga\"{e}l Gu\'eh\'eneuc is with the Department of Computer Science and Software Engineering, Concordia University, Canada\protect\\
E-mail: yann-gael.gueheneuc@concordia.ca
}%
}



\maketitle

\input{Sections/0-abstract}


 \input{Sections/1-introduction}
 \input{Sections/2-relatedWorks}
 \input{Sections/3-methodology}
 \input{Sections/4-RQ1}
\input{Sections/5-RQ2}
\input{Sections/6-RQ3}
\input{Sections/7-RQ4}
\input{Sections/8-RQ5}
\input{Sections/9-discussion}
\input{Sections/9.1-Threats2Validity}
\input{Sections/10-conclusion}

\section*{Replication Package}
The replication package can be accessed using the following DOI: \href{https://doi.org/10.5281/zenodo.15723658}{10.5281/zenodo.15723658}

\section*{Acknowledgments}
The Canada Research Chair program partly funded this work.

 




\bibliographystyle{IEEEtran}
\bibliography{ref}
\bibliographystyleP{IEEEtran}
\bibliographyP{studies}

\input{Sections/appendices}
\end{document}

%% file: Sections/0-abstract.tex
\begin{abstract}
Scalability and maintainability challenges in monolithic systems have led to the adoption of microservices, which divide systems into smaller, independent services. However, migrating existing monolithic systems to microservices is a complex and resource-intensive task, which can benefit from machine learning (ML) to automate some of its phases. Choosing the right ML approach for migration remains challenging for practitioners. 
Previous works studied separately the objectives, artifacts, techniques, tools, and benefits and challenges of migrating monolithic systems to microservices. No work has yet investigated systematically existing ML approaches for this migration to understand the \revised{automated migration phases}, inputs used, ML techniques applied, evaluation processes followed, and challenges encountered.
We present a systematic literature review (SLR) that aggregates, synthesises, and discusses the approaches and results of 81 primary studies (PSs) published between 2015 and 2024. 
We followed the Preferred Reporting Items for Systematic Review and Meta-Analysis (PRISMA) statement to report our findings and answer our research questions (RQs). We extract and analyse data from these PSs to answer our RQs. We synthesise the findings in the form of a classification that shows the usage of ML techniques in migrating monolithic systems to microservices.
The findings reveal that some phases of the migration process, such as monitoring and service identification, are well-studied, while others, like packaging microservices, remain unexplored. Additionally, the findings highlight key challenges, including limited data availability, scalability and complexity constraints, insufficient tool support, and the absence of standardized benchmarking, emphasizing the need for more holistic solutions. 

\end{abstract}

%% file: Sections/1-introduction.tex
\section{Introduction}
\label{Section: introduction}

As software systems evolve, they often grow in size and complexity due to the continuous addition of new functionalities. This growth frequently results in tightly coupled components with diminished cohesion, leading to architectural and operational challenges that impede scalability, maintainability, and deployment.\cite{ali2018architecture}. Traditionally, such systems have been built using monolithic architectures, where functionality is centralised within large, interconnected components \cite{dragoni2017microservices}. While this architecture was widely adopted in the past, the increasing demand for flexibility and faster release cycles has revealed its inherent limitations \cite{fritzsch2019microservices}.

In response to these limitations, microservice architecture has emerged as a solution to divide monolithic systems into or build new systems as smaller, independent services \cite{tapia2020monolithic}. This division enables improved scalability, faster deployment, and more efficient maintenance, making microservices an attractive alternative for modern software development and evolution \cite{abgaz2023decomposition}.

However, migrating existing monolithic systems to a microservices architecture remains a complex and resource-intensive task \cite{sarkar2018towards}. It involves identifying service boundaries and packaging them into self-contained microservices with defined APIs. Effective migration further requires robust strategies for deploying microservices and guaranteeing their scalability, security, and fault tolerance. Additionally, continuous monitoring is essential to ensure optimal functionality and maintain the system's operational health.

While many approaches have been proposed to address the various phases of migrating monolithic systems to microservices, such as heuristic-based methods and domain-driven design, these techniques often require manual intervention and lack adaptability across different systems\cite{abdellatif2021taxonomy}. Recent advancements in machine learning (ML) offer a promising avenue to tackle these challenges, including identifying service boundaries, analysing interdependencies, and predicting failure. ML techniques excel at processing large and complex datasets, uncovering patterns, and supporting decision-making processes that are otherwise difficult to achieve with traditional approaches. For example, ML can assist in automating repetitive and error-prone tasks, such as clustering related components or optimising deployment strategies, potentially reducing migration time and improving accuracy and consistency \cite{toumi2023machine}. Although business process (BP) analysis and modernisation are often crucial first steps in practice, most ML-based studies focus on migration tasks involving source code or system architecture. A few works leverage business process models as input, but typically assume these models are already available. The automation of BP discovery and refactoring prior to migration remains largely unexplored and presents a valuable opportunity for future research using NLP, process mining, or large language models.

This systematic literature review (SLR) aims to explore the role of ML in this domain by addressing key research questions, such as the phases of migration it automates, the characteristics of the input data it leverages, the types of ML approaches applied, how these approaches are evaluated, and the challenges faced by researchers. By answering these questions, this study seeks to provide a comprehensive understanding of the current state of ML-based migration techniques and identify opportunities for future research.

Overall, the findings suggest that while certain phases of the migration process from monolithic systems to microservices, such as monitoring and service identification, have been extensively studied, other phases remain in their early stages of exploration \cite{oumoussa_Access_2024_evolution}. Notably, there is a significant gap in approaches specifically designed for the later stages of the migration process, including packaging microservices, generating the necessary code for microservice APIs, and implementing design patterns critical to their operation. Despite the progress made in some phases like identification, deployment and monitoring, other phases like pre-migration and packaging are not treated. We also found that the current approaches face several challenges that hinder their adoption in practical scenarios. These challenges include the insufficient availability of high-quality data, which is essential for training reliable machine learning models, as well as scalability and complexity concerns that restrict the applicability of these methods to large-scale and real-world systems \cite{fritzsch_ieee_2019_microservices}. Furthermore, many existing techniques lack tool support, making them difficult to implement effectively in the industry. The absence of standardised metrics, datasets, and baselines further complicates the evaluation and comparison of these approaches, limiting the ability to measure progress and identify best practices \cite{ahmed2023deep}.

While several studies have explored specific aspects of microservice migration and the associated challenges, to the best of our knowledge, no prior work investigated systematically existing ML approaches for this migration to understand the automated phases, used inputs, applied ML techniques, followed evaluation processes, and encountered challenges.

\revised{Motivated by this gap, the goal of this study is to understand how machine learning is used to migrate monolithic systems to microservices. This leads to our main research question:  \\
\textit{How is machine learning used to migrate monolithic systems to microservices?}  }

From this main question, we derive the following research questions (RQs):

\begin{enumerate}
\item[\ding{202}] \textbf{RQ1}:\rqone
\item[\ding{203}] \textbf{RQ2}:\rqtwo
\item[\ding{204}] \textbf{RQ3}:\rqthree
\item[\ding{205}] \textbf{RQ4}:\rqfour
\item[\ding{206}] \textbf{RQ5}:\rqfive
\end{enumerate}

To answer these research questions, we followed the updated Preferred Reporting Items for Systematic Review and Meta-Analysis (PRISMA) statement for reporting systematic reviews~\cite{2021prisma}. We screened 2,301 potentially relevant studies from eight digital libraries, considering publications between 2015 and 2024. Using inclusion and exclusion criteria along with snowball sampling, we assessed the quality of the primary studies based on their design, methodology, analysis, and conclusions. We retained a total of 81 PSs for analysis.

The primary contribution of this study is the development of a comprehensive understanding of how ML can be used to support the migration of monolithic systems to microservices. This is achieved through the following key contributions:

\begin{enumerate}
\item[\ding{202}] A systematic analysis of the migration phases automated using ML techniques, identifying gaps in automation across the migration lifecycle (RQ1).

\item[\ding{203}] An organised synthesis of the types of inputs used in ML-driven migration, including their granularity, sources and how it is preprocessed, forming a basis for understanding the input usage in migration (RQ2).

\item[\ding{204}] A detailed analysis of machine learning techniques employed for migration, highlighting commonly used models, emerging techniques, and their features (RQ3).

\item[\ding{205}] An exploration of evaluation practices for ML-based migration approaches, identifying common metrics, benchmarking, and success criteria (RQ4).

\item[\ding{206}] A discussion of the challenges encountered when applying ML for migration, including scalability, data availability, and the interpretability of models, along with directions for future research (RQ5).
\item[\ding{207}] A set of recommendations for practitioners and researchers, derived from the insights gathered, to help address the challenges of applying ML techniques during monolithic-to-microservices migration.

\end{enumerate}

We organise the remainder of this paper as follows. Section \ref{Section: relatedWorks} presents the background and related work on monolithic to microservices migration. Section \ref{Section: Methods} describes the SLR methodology used and provides an overview of the analysis process. Section \ref{Section: rq1} focuses on phases of migration that are currently automated using ML. Section \ref{Section: rq2} examines the types of data collected and processed as inputs for ML during migration. Section \ref{Section: rq3} identifies the ML techniques applied in this context. Section \ref{Section: rq4} discusses how ML-based approaches are evaluated for migration. Section \ref{Section: rq5} investigates the challenges and limitations of applying ML to migration. Section \ref{Section: Discussion} discusses the observations and recommendations. Section \ref{sec:Threats2Validity} outlines the potential threats to the validity of our study. Finally, Section \ref{Section: Conclusion} summarises the findings and suggests directions for future research.

%% file: Sections/2-relatedWorks.tex
\section{Related Work}
\label{Section: relatedWorks}

The migration of monolithic applications to microservices has gained significant attention in recent years due to its potential to improve scalability, flexibility, and maintainability in software systems. Several studies have explored various aspects of this migration, ranging from decomposition frameworks to the challenges faced during the migration. This section presents an overview of the most relevant works, focusing on decomposition frameworks, migration strategies, and the challenges associated with microservice adoption.

\begin{table*}[ht]
\centering
\caption{Closely Related Works}
\label{tab:related_work}
\footnotesize
\scalebox{0.83}{
\begin{tabular}[t]{p{0.02\linewidth}p{0.16\linewidth}p{0.04\linewidth}p{0.04\linewidth}p{0.76\linewidth}}
\toprule
\textbf{Study} &
  \textbf{Focus} &
  \textbf{Year} &
  \textbf{Method} &
  \textbf{RQs} \\ \midrule
 \cite{related_moreschini2025ai} & Use of AI techniques for improving any quality attribute of microservices during the DevOps phases & 2025&SMS&
  \begin{minipage}[t]{1.0\linewidth} 
\begin{itemize}[noitemsep,topsep=0pt,partopsep=0pt,leftmargin=*]
\item In which industry domains is AI used for microservices?
\item What challenges arise in applying AI to microservices design?   
\item What benefits does AI offer over traditional microservices design?
\item What are the emerging AI trends in microservices design?
\item What input artifacts are used in AI-driven microservices design?
\end{itemize}
\end{minipage}
\\ \midrule
  
\cite{related_narvaez2025designing} & AI for improving microservices design, decomposition, and validation & 2025&SLR&
  \begin{minipage}[t]{1.0\linewidth} 
\begin{itemize}[noitemsep,topsep=0pt,partopsep=0pt,leftmargin=*]
\item What AI methods support microservices design in new software systems?
\item What challenges arise in applying AI to microservices design?   
\item What benefits does AI offer over traditional microservices design?
\item What are the emerging AI trends in microservices design?
\item What input artifacts are used in AI-driven microservices design?
\end{itemize}
\end{minipage}
\\ \midrule
\cite{saucedo_TSE_2024_migration} &
Microservice: identification techniques, tools, factors, issues, and migration benefits&
2024 &
SMS &
\begin{minipage}[t]{1.0\linewidth} 
\begin{itemize}[noitemsep,topsep=0pt,partopsep=0pt,leftmargin=*]
\item How are monolithic systems to microservices migration done?  
\item Which factors have caused the migration of monolithic systems to microservices?    
\item What are the issues and benefits of migrating towards microservices?
\end{itemize}
\end{minipage}
\\ \midrule
\cite{oumoussa_Access_2024_evolution} &
  Microservice: Evolution &
  2024 &
  SLR &
 \begin{minipage}[t]{1.0\linewidth} 
\begin{itemize}[noitemsep,topsep=0pt,partopsep=0pt,leftmargin=*]
\item How has the area of study on microservices identification evolved? 
\item What is the current state-of-the-art in microservices identification research?
\item What are the current and potential challenges associated with microservices identification? 
\end{itemize}
\end{minipage}
\\ \midrule
\cite{saucedo_elsevier_2024_migration} &
  Microservice: Migration Techniques, Tools, Factors,   and Benefits &
  2024 &
  SMS &
 \begin{minipage}[t]{1.0\linewidth} 
\begin{itemize}[noitemsep,topsep=0pt,partopsep=0pt,leftmargin=*]
\item How are monolithic systems to microservices migration done?
\item Which factors have caused the migration of monolithic systems to microservices?
\item What are the issues and benefits of migrating towards microservices?
\end{itemize}
\end{minipage}
\\ \midrule
\cite{razzaq_wiley_2023_systematic} &
  Migration Approaches,   Challenges, Successful factors, and    Potential for industrial adoption &
  2022 &
  SMS &
  \begin{minipage}[t]{1.0\linewidth} 
\begin{itemize}[noitemsep,topsep=0pt,partopsep=0pt,leftmargin=*]
\item How many articles in this field of research can be found each year? \item What are the primary venues for the production and printing of the research?\item What are the principal publication types in the research area?
\end{itemize}
\end{minipage}
\\ \midrule
\cite{Abgaz_TSE_2023_decomposition} &
  Microservices:   Decomposition &
  2023 &
  SLR &
 \begin{minipage}[t]{1.0\linewidth} 
\begin{itemize}[noitemsep,topsep=0pt,partopsep=0pt,leftmargin=*]
\item What are the existing approaches, tools, and methods observed in the decomposition of monolith to microservices?
\item What are the metrics, datasets, and benchmarks used for evaluating monolith decomposition to microservices?
\item What research gaps can be identified in the current literature?
\end{itemize}
\end{minipage}
\\ \midrule
\cite{velepucha_ICI2ST_2021_monoliths} &
  Microservices: Quality Assurance &
  2021 &
  SLR &
\begin{minipage}[t]{1.0\linewidth} 
\begin{itemize}[noitemsep,topsep=0pt,partopsep=0pt,leftmargin=*]
\item What problems and challenges are there for the migration process of monolithic applications to microservices? 
\end{itemize}
\end{minipage}
  \\ \midrule
\cite{bushong_MDPI_2021_microservice} &
  Microservice: Evolution &
  2021 &
  SMS &
 \begin{minipage}[t]{1.0\linewidth} 
\begin{itemize}[noitemsep,topsep=0pt,partopsep=0pt,leftmargin=*]
\item What methods and techniques are used in microservice analysis?
\item What are the problems or opportunities that are addressed using microservice analysis techniques?    
\item Does microservice analysis overlap with other areas of software analysis,   or are new methods or paradigms needed?
\item What potential future research directions are open in the area of microservice analysis?\end{itemize} 
\end{minipage}
\\ \midrule
\cite{ponce_IEEE_2019_migrating} &
  Microservice: Migration Techniques &
  2019 &
  RR &
\begin{minipage}[t]{1.0\linewidth} 
\begin{itemize}[noitemsep,topsep=0pt,partopsep=0pt,leftmargin=*]
\item What are the migration techniques proposed in the literature?
\item In what types of systems have the proposed techniques been applied?
\item What type of validation do the authors of the techniques use?
\item Are there challenges associated with migration from monolith to microservices?
\end{itemize} 
\end{minipage}
\\ \midrule
\cite{fritzsch_springer_2019_monolith} &
  Microservice: Refactoring Approaches &
  2019 &
  SLR &
   \begin{minipage}[t]{1.0\linewidth} 
\begin{itemize}[noitemsep,topsep=0pt,partopsep=0pt,leftmargin=*]
\item What are existing architectural refactoring approaches in the context of decomposing a monolithic application architecture into Microservices, and how can they be classified with regard to the techniques and strategies used?
\end{itemize}
\end{minipage}
\\ \midrule
\cite{silva_springer_2019_strategies} &
  Microservice: Migration Strategies &
  2019 &
  SLR &
 \begin{minipage}[t]{1.0\linewidth} 
\begin{itemize}[noitemsep,topsep=0pt,partopsep=0pt,leftmargin=*]
\item Which strategies have been reported in the literature to support the migration of legacy software systems to microservices-based architecture?
\item Which lessons learned have been reported in the literature regarding challenges and advantages perceived as a consequence of the aforementioned migration?
\end{itemize}
\end{minipage} 
\\ \midrule

\cite{fritzsch_ieee_2019_microservices} &
  Microservice: Migration Intentions, Strategies, and   Challenges &
  2019 &
  IS &
 \begin{minipage}[t]{1.0\linewidth} 
\begin{itemize}[noitemsep,topsep=0pt,partopsep=0pt,leftmargin=*]
\item What are the intentions for migrating existing systems to Microservices? \item Which Microservices migration strategies and decomposition approaches do companies apply? \item What are the major technical and organisational challenges during a Microservices migration?
\end{itemize}
\end{minipage}
\\ \midrule
\cite{mparmpoutis_2022_using} &
  Microservice: Migration Artefacts &
  2023 &
  SMS &
 \begin{minipage}[t]{1.0\linewidth} 
\begin{itemize}[noitemsep,topsep=0pt,partopsep=0pt,leftmargin=*]
\item During the process of re-architecting a legacy system, are data-driven artifacts like database schema and the state of data used for identifying potential microservices?
\item How are data-driven artifacts, like the database, used in the process of software migration to services/microservices-based architecture?
\end{itemize}
\end{minipage}
\\ \midrule
\cite{di_IEEE_2018_migrating} &
  Microservice: Migration practices &
  2018 &
  IS &
  \begin{minipage}[t]{1.0\linewidth} 
\begin{itemize}[noitemsep,topsep=0pt,partopsep=0pt,leftmargin=*]
\item What are the activities carried out by practitioners when migrating towards a microservice-based architecture? \item What are the challenges faced by practitioners when migrating towards a microservice architecture? 
\end{itemize}
\end{minipage}
\\ \midrule
\cite{kalske_Springer_2018_challenges} &
  Microservice: Migration Reasons and Challenges &
  2018 &
  IS &
  \begin{minipage}[t]{1.0\linewidth} 
\begin{itemize}[noitemsep,topsep=0pt,partopsep=0pt,leftmargin=*]
\item NP \end{itemize}\end{minipage} \\ \midrule
\cite{velepucha_IEEE_2021_monoliths} &
  Microservice: Migration Problems and Challenges &
  2021 &
  SMS &
   \begin{minipage}[t]{1.0\linewidth} 
\begin{itemize}[noitemsep,topsep=0pt,partopsep=0pt,leftmargin=*]
\item NP \end{itemize}\end{minipage} \\ \midrule
\cite{capuano_IEEE_2022_systematic} &
  Microservice: Migration Quality Attributes &
  2022 &
  SLR &
 \begin{minipage}[t]{1.0\linewidth} 
\begin{itemize}[noitemsep,topsep=0pt,partopsep=0pt,leftmargin=*]
\item  Which studies implement a quality-driven approach to migrate to microservices? \item Which are the quality attributes analyzed in the migration phases?\item In which migration phase is the quality-driven process implemented?
\end{itemize}
\end{minipage}
\\ \bottomrule
\end{tabular}
}
\begin{tablenotes}
\centering
\small
    \item[\ding{90}] \textbf{IS}: Industry Survey; \textbf{SLR}: Systematic Literature Review; \textbf{SMS}: Systematic Mapping Study; \textbf{RR}: Rapid Review; \textbf{NP}: Not Provided.
\end{tablenotes}

\end{table*}

Table \ref{tab:related_work} summarises the existing studies closely related to this SLR. Following established frameworks in the literature~\cite{Abgaz_TSE_2023_decomposition, fritzsch_springer_2019_monolith, saucedo_elsevier_2024_migration}, the migration process considered in this study is structured into five phases: Pre-migration, Identification, Packaging, Deployment, and Monitoring. These phases result from a synthesis and refinement of prior works, tailored to the scope of ML-supported migration. This structuring aims to capture the full lifecycle of monolith-to-microservices transition while highlighting phases where automation through machine learning could provide concrete benefits.

Abgaz et al.\cite{Abgaz_TSE_2023_decomposition} introduced the Monolith to Microservices Decomposition Framework (M2MDF), identifying major phases and key elements involved in the decomposition process. They analyzed existing methods, tools, and metrics used for decomposition, proposing future directions for refining techniques. Fritzsch et al.\cite{fritzsch_springer_2019_monolith} similarly classified refactoring approaches for decomposition. However, these studies do not consider the role of machine learning (ML) in supporting or automating migration tasks.
Migration challenges and quality attributes have also been addressed. Velepucha et al.\cite{velepucha_ICI2ST_2021_monoliths} identified quality improvement tactics, while Ponce et al.\cite{ponce_IEEE_2019_migrating} categorised migration techniques by automation and validation. Razzaq et al.\cite{razzaq_wiley_2023_systematic} examined organizational factors influencing successful migration. Bushong et al.\cite{bushong_MDPI_2021_microservice} explored overlaps between traditional and microservice-specific software analysis. While insightful, these works do not explore ML-driven migration strategies, inputs, or evaluation methods.

Saucedo et al.~\cite{saucedo_elsevier_2024_migration} proposed a structured migration process based on real-world industrial cases, offering a catalogue of tools and influencing factors. However, their work does not examine ML-based automation. Daniel et al.\cite{related_narvaez2025designing} discussed microservices design inputs and challenges, while Sergio\cite{related_moreschini2025ai} mapped AI techniques in DevOps but did not focus on migration-specific ML strategies.

Other studies reviewed migration strategies and lessons learned. Silva et al.\cite{silva_springer_2019_strategies} and Capuano et al.\cite{capuano_IEEE_2022_systematic} highlighted strategic challenges, whereas Kalske et al.\cite{kalske_Springer_2018_challenges} focused on organizational motivations. Mparmpoutis et al.\cite{mparmpoutis_2022_using} leveraged data artifacts for service identification, and Kazanavivcius et al.\cite{kazanavivcius_IEEE_2019_migrating} examined legacy migration drawbacks. None of these works, however, evaluated ML-based techniques, their automation potential, or supporting inputs and metrics.
Di Francesco et al.\cite{di_IEEE_2018_migrating} studied practitioner challenges during migration, and Fritzsch et al.\cite{fritzsch_ieee_2019_microservices} explored migration intentions and organizational factors. Yet, these studies do not consider ML or its evaluation in migration automation.

In summary, existing works provide valuable foundations for understanding microservice migration challenges, strategies, and decomposition methods. However, to our knowledge, no prior study has systematically investigated the use of ML techniques in this context. Our study addresses this gap by examining the phases of migration automated using ML, the input types, techniques applied, evaluation practices, and challenges faced.

%% file: Sections/3-methodology.tex
\section{Research Method}
\label{Section: Methods}
We followed the updated PRISMA guidelines \cite{page2021prisma, page2021prismae} and Kitchenham et al.  \cite{kitchenham2022segress} guidelines to review and report our findings.
We used three main phases: planning, conducting, and reporting the review. During the planning phase, we defined the objective of SLR and reviewed the protocol. The objective of this SLR is defined in Section \ref{Section: introduction}. This section defines the review protocol for conducting the SLR.
It consists of six steps: \begin{enumerate*} \item[\ding{172}]defining the research questions, \item[\ding{173}] formulating the search query, \item[\ding{174}] selecting the studies, \item[\ding{175}] snowballing, \item[\ding{176}] assessing the quality of the studies, and \item[\ding{177}]extracting and analysing the data. \end{enumerate*} 
In the following subsections, we explain each step of our review protocol.

\subsection{Research Questions (RQs)}

This study answers the following RQs:

\textbf{RQ1}:\rqone  \par
\textbf{Rationale:}
By analysing the phases and the tasks supported by ML, we want to understand the role and impact of ML in the migration process. This investigation provides insights into how ML is integrated into current migration approaches. 

 \textbf{RQ2}:\rqtwo \par
\textbf{Rationale:}
By categorising inputs based on their types, granularities, sources, and preprocessing methods, we want to create a structured classification of the inputs and understand how inputs are gathered, prepared, and utilised in machine learning models for microservices migration.

\textbf{RQ3}:\rqthree \par
\textbf{Rationale:}
By investigating the ML approaches in the migration, we want to (1) identify and classify the ML models used, (2) examine their learning paradigms, and (3) analyse the features leveraged to enhance model performance. This RQ provides a comprehensive understanding of how ML contributes to different phases of the migration.

 \textbf{RQ4}:\rqfour \par
\textbf{Rationale:}
By investigating how ML-based methods for microservices migration are evaluated, we want to (1) identify the evaluation metrics used to assess the quality and performance of these methods, (2) explore the benchmarks employed for comparative analyses, (3) understand the success criteria that define their effectiveness, and (4) analyse the tools available for implementing and evaluating these approaches.  

\textbf{RQ5}:\rqfive\par
\textbf{Rationale:} 
By examining the limitations, we aim to identify and analyse the key challenges associated with applying machine learning techniques in the migration from monolithic architectures to microservices.

\subsection{Search Query}
\label{subsec:searchstrategy}

We formulated our search query by applying the PICO (Population, Intervention, Comparison, Outcome) framework \cite{pico}. We followed the following steps:
\begin{itemize}
\item[\ding{202}] Obtaining the main terms from our main research question, as stated in the introduction. 
\item[\ding{203}] Identifying the possible synonyms of the main terms.
\item[\ding{204}] Applying the Boolean OR to combine possible synonyms of the main terms.
\item[\ding{205}] Applying the Boolean AND to combine expressions in the previous step.
\end{itemize}
As a result of PICO framework, we formulated the following search query:
\begin{boxD}
(Monolith* OR Exist* OR Legac*) 
AND 
(Microservice* OR Micro-service* OR MSA)
AND
(Migrat* OR Identif* OR Decompos* OR Extract* OR Transform* OR Refactor* OR Transit* OR Creat* OR Genera*)
AND (Machine learning OR ML 
OR Neural Network* OR *coder 
OR *supervised 
OR Reinforcement Learning 
OR Model*) 
\end{boxD}
To obtain more comprehensive results, we used the asterisk (*) in search queries as a wildcard to match any sequence of characters.

\subsection{Studies Selection}
\begin{figure}[ht]
  \includegraphics[width=\linewidth]{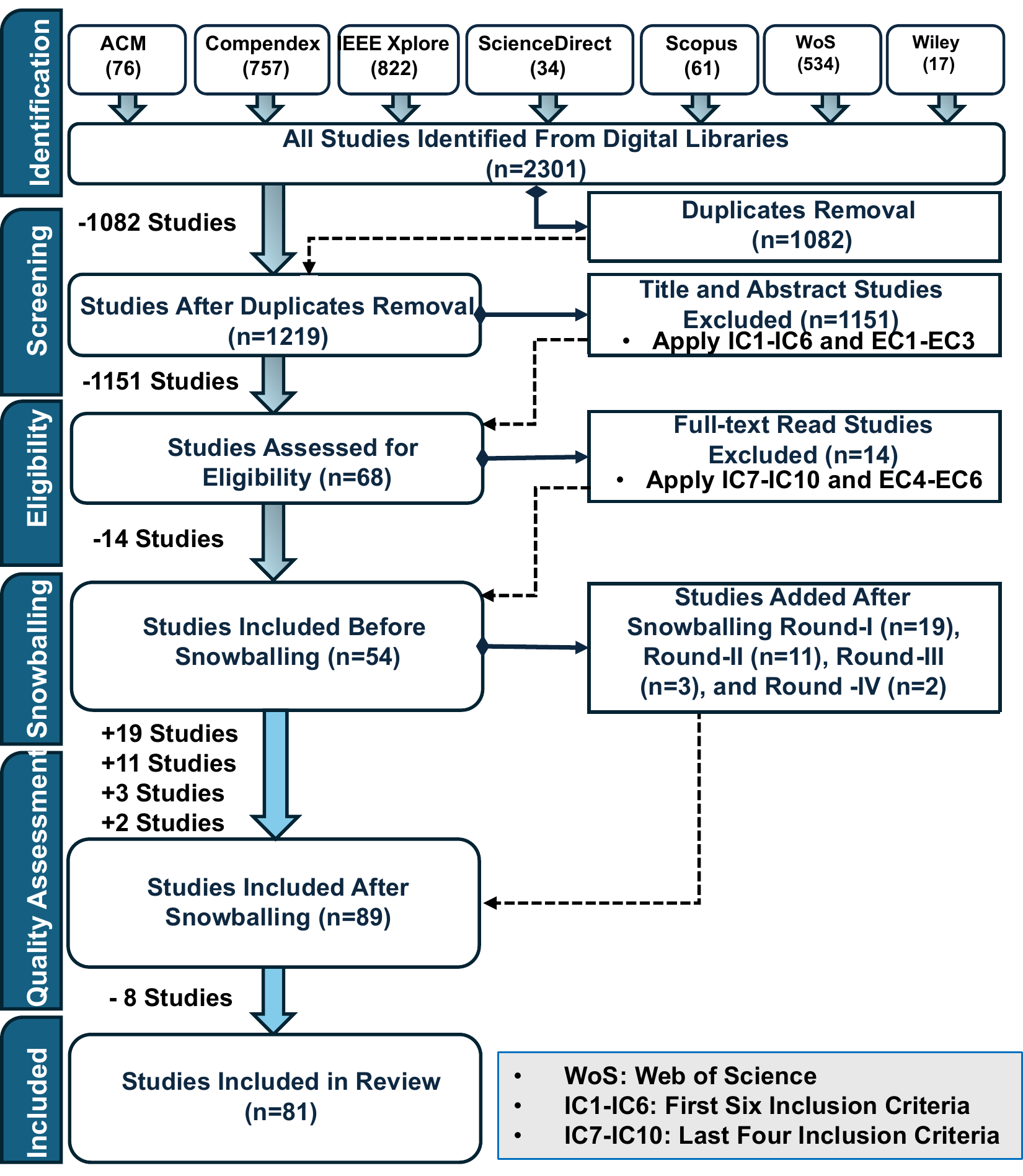}
  \caption{PRISMA Flow for Primary Studies Selection}
  \label{fig:prisma_overview}
\end{figure}

We applied the PRISMA steps to select the PSs. The main steps include database identification, removal of duplicates, screening, eligibility assessment, multiple rounds of both \revised{backwards} and forward snowballing, and quality assessment of each study. Figure \ref{fig:prisma_overview} summarises those steps.
\subsubsection{Databases Identification}\label{sec:db_identification}
We selected seven online digital libraries: ACM Digital Library, Compendex, IEEE Xplore, ScienceDirect, SpringerLink, Scopus, Web of Science, and Wiley. These libraries are widely used for literature reviews in software engineering, as recommended by Dyba et al.~\cite{dyba2007applying}. 

We applied our search query to each of these digital libraries. However, some libraries impose restrictions when performing queries. For instance, ScienceDirect limits queries to a maximum of eight connectors, while the ACM Digital Library does not allow wildcards. We adjusted the search query to meet the specific requirements of each library. Table  \ref{tab:dbsearchresults} shows the search queries executed in each digital library and the number of studies retrieved.
\begin{table}[ht]
\caption{Database Search Results}
\label{tab:dbsearchresults}
\centering
\footnotesize
\resizebox{0.95\columnwidth}{!}{%
\begin{tabular}{@{}lrr@{}}
\toprule
\textbf{Database} & \textbf{All   Search Studies} & \textbf{Selected   Studies (PSs)} \\ \midrule
ACM              & 76  & 10 \\
Compendex        & 757 & 21 \\
IEEE   Xplore    & 822 & 37 \\
ScienceDirect    & 34  & 3 \\
Scopus           & 61  & 6 \\
Web   of Science & 534 & 1 \\
Wiley            & 17  & 3 \\ \bottomrule
\textbf{Total}           & \textbf{2301}                 & \textbf{81}                        \\ \bottomrule
\end{tabular}%
}
\end{table}
Our search was confined to English-language, peer-reviewed scholarly articles published in journals, conferences, and workshops between 2015 and 2024. This time frame was chosen due to the increase in microservices-related literature starting in 2015 \cite{vural2017systematic}. We initially retrieved a total of 2301 studies from seven libraries.

\subsubsection{Duplicates Removal}
\label{sec:de_duplication}
Duplicates were identified based on an exact match of the study's title, first author, and venue (conference or journal). This reduced the total from 2301 studies to 1219.

\subsubsection{Screening}
\label{sec:screening}
We defined inclusion and exclusion criteria and applied them to select relevant primary studies while excluding irrelevant ones.

\textbf{Inclusion Criteria:} We considered the following inclusion criteria for PSs selection:
\begin{itemize}
\item \textbf{IC1:} The study is written in English.
\item \textbf{IC2:} The study is published between 2015 and 2024.
\item \textbf{IC3:} The study is published in journals, conferences, or workshops. 
\item \textbf{IC4:} The study has at least 4 pages.
\item \textbf{IC5:} The study focuses on the migration of monolith applications into microservices.
\item \textbf{IC6:} The study uses a machine/deep learning algorithm.
 \item \textbf{IC7:} The study provides enough information to answer at least three research questions (RQs).
 \item \textbf{IC8:} The study has its full text available online.
\item \textbf{IC9:} The study provides sufficient migration details.
\item\textbf{IC10:} The study uses an automated or semi-automated migration approach.

\end{itemize}  
\textbf{Exclusion Criteria:} We considered the following exclusion criteria:
\begin{itemize}
\item\textbf{EC1:} The study is a secondary source (e.g., literature review, survey, opinion piece, or discussion).
\item\textbf{EC2:} The study has not been peer-reviewed.
\item\textbf{EC3:} The study is a graduate thesis or project report.
\item\textbf{EC4:} The study's full text is not available online.
\item\textbf{EC5:} The study does not provide enough details.
\item\textbf{EC6:} The study does not provide an automated or semi-automated migration approach.
\end{itemize}

We applied our inclusion and exclusion criteria in two \revised{steps: first}, using the titles and abstracts, then the full texts. Criteria IC1-IC6 and EC1-EC3 were applied to the titles and abstracts, while IC7-IC10 and EC4-EC6 were applied to the full texts. We considered that a study provided 'sufficient migration details' (IC9) if it described at least the system analyzed, the ML technique applied, and the evaluation procedure used. Conversely, studies lacking two of these three aspects were excluded under EC5. Additionally, EC5 encompassed studies that solely described theoretical approaches without providing experimental validation or application to real systems or datasets, in order to ensure the practical relevance of the selected primary studies. We acknowledge that some inclusion and exclusion criteria are logically complementary (e.g., IC8/EC4, IC9/EC5 and IC10/EC6) describe opposing conditions. Additionally, IC7 partially overlaps with EC5. We explicitly state both inclusion and exclusion criteria separately to comply with the PRISMA~\cite{2021prisma} guidelines, which recommend clearly articulating both positive and negative selection criteria to enhance transparency and reproducibility. To clarify the practical application of the criteria, we present in Table \ref{tab:exclusion-examples} (in Appendix ) examples of excluded studies for IC7–IC10, EC5, and EC6, along with brief justifications for their exclusion.
During the initial screening, we assessed the titles, abstracts, and page counts, and determined if the studies qualified as primary. Two authors independently conducted the screening using predefined inclusion and exclusion criteria. To ensure consistency, we compared the screening results for 50 randomly selected studies using Cohen's Kappa \cite{kepacohen1968weighted} 

We calculated Cohen's Kappa, achieving near-perfect agreement (k=0.85), highlighting the consistency of our screening. For the remainder of the process, authors met regularly to review results and resolved disagreements through discussion and consensus.
After completing the initial screening, we selected 68 studies for full-text review as potential PSs.
\subsubsection{Eligibility Assessment}
\label{sec:eligibility_assessment} 
In the second round of screening, two authors independently applied IC7-IC10 to 68 studies by thoroughly reviewing them. To ensure a shared understanding of the inclusion and exclusion criteria, we compared the results of 15 randomly selected studies using Cohen's Kappa. We calculated a near-perfect agreement (k=0.93), demonstrating strong consistency between the two authors. This allowed us to confidently proceed with the eligibility assessment for the remaining studies. Ultimately, we identified 54 PSs that met the eligibility criteria.

\subsection{Snowballing}
\label{sec:snowballing}

We conducted four rounds of both forward and backwards snowballing to identify additional primary studies (PSs). In Round 1, we reviewed the references and citations of 54 initial PSs and identified 1,347 potential studies. After removing duplicates, we retained 1,172 studies and applied our inclusion and exclusion criteria, which resulted in 19 new PSs, bringing the total to 73. In Round 2, we used the 19 newly added PSs to identify 602 potential studies and selected 11 new PSs using the same criteria, raising the total to 84. In Round 3, we reviewed the references and citations of the 11 new PSs, found 190 potential studies, and selected 3 additional PSs, bringing the total to 87. In Round 4, we examined the references and citations of the 3 PSs from the previous round, identified 43 potential studies, and selected 2 more PSs. This brought the final total to 89 PSs.
Table \ref{tab:snowballing} presents the details from each round.
\begin{table}[ht]
\caption{Snowballing Results}
\label{tab:snowballing}
\resizebox{\columnwidth}{!}{%
\begin{tabular}{@{}lrrrr@{}}
\toprule
\textbf{Snowballing} & \textbf{Round} & \textbf{Retrieved} & \textbf{No duplicates} & \textbf{Included} \\ \midrule
Backward \& Forward & 1 & 1346 & 1172 & 19 \\
Backward \& Forward & 2 & 602  & 375  & 11 \\
Backward \& Forward & 3 & 190  & 135  & 3  \\
Backward \& Forward & 4 & 43   & 43   & 2  \\ \midrule
\textbf{Total}       & \textbf{}      & \textbf{}          & \textbf{1725}                  & \textbf{35}       \\ \bottomrule
\end{tabular}%
}
\end{table}

\subsection{Quality Assessment}
 Following the guidelines proposed by Li et al. \cite{li_Elsevier_2021_understanding} and Dyba et al. \cite{dyba_Elsevier_2008_Empirical}, we developed a quality assessment checklist comprising eight questions (or quality criteria) to evaluate the quality of our primary studies (PSs). Table \ref{tab:qualityCriteria} shows the quality criterias we used in this study based on Li et al.'s work \cite{li_Elsevier_2021_understanding}.
\begin{table*}[ht]
\caption{Quality Criteria extracted from \cite{li_Elsevier_2021_understanding}}
\label{tab:qualityCriteria}
\centering
\begin{threeparttable}
\footnotesize
\scalebox{0.95}{
\begin{tabularx}{\linewidth}[t]{
    >{\hsize=0.05\hsize}X
    >{\hsize=1.95\hsize}X 
  }
\toprule
\textbf{No} &
  \textbf{Quality Criteria} \\ \midrule
Q1 &
  Is there a clear statement of the aims of the research?   Consider: Is there a rationale for why the study was undertaken? \\
Q2 &
  Was the research design appropriate to address the aims of the research? Consider: Did the researcher justify the design of the research? \\
Q3 &
  Was the research method implemented in a way that addressed the research issue? Consider: Has the researcher discussed the process or the details of the methods that were chosen/proposed? \\
Q4 &
  Is there a clear statement of findings? Consider: Has an adequate discussion/evaluation of the evidence identified or method proposed,   both for and against the researchers' arguments, been demonstrated? \\
Q5 &
  Has the limitation or future work been considered adequately? Consider:  Did the researcher examine the limitations and future work? \\
Q6 &
  Is there an adequate description of the context in which the research was carried out? Consider: Did the researcher explain how the context influenced the study, and was this context relevant to the aim of the research? \\
Q7 &
  Was the data collected in a way that addressed the research issue? Consider: Did the researcher collect data from relevant sources, and was the data collection process clearly described and ethically sound? \\
Q8 &
  Was the data analysis sufficiently rigorous? Consider: Did the researcher provide a thorough and unbiased examination of the data, and were the limitations of the analysis acknowledged? \\ \bottomrule
\end{tabularx}%
}
\end{threeparttable}
\end{table*}
Responses to each question were: "Yes" (1 point), "Partially" (0.5 points), or "No" (0 points). Three authors independently applied this checklist to each study, discussed any discrepancies, and achieved consensus on the scores.
We calculated the quality score for each study by summing the individual scores and converting this total into a percentage. Studies achieving an 80\% score or higher were retained. Using this criterion, we excluded eight studies falling below the threshold and included 81 PSs in our review.

\subsection{Data Extraction and Analysis}
We analyzed and extracted the necessary data from each primary study (PS) to address our research questions (RQ1-RQ5). Table \ref{tab:dataTemplate} presents the key data items extracted, including the item name, a brief description, and the corresponding research question it addresses. The full details of the extracted data are available in our publicly accessible replication package.

 To ensure accuracy and consistency, two researchers independently extracted data using a shared Excel template with dedicated columns for each data field. Initially, each researcher processed 50\% of the PSs. This was followed by a full cross-validation phase, during which each extraction was reviewed by the other researcher. We achieved strong inter-rater reliability, with only 8\% of the data points requiring arbitration. In those cases, a senior third researcher resolved discrepancies by re-examining the original source papers and documenting the final decision.

\begin{table*}[ht]
\caption{ Data Extraction Template}
\centering
\label{tab:dataTemplate}
\begin{threeparttable}
\footnotesize
\scalebox{0.85}{
\begin{tabularx}{\linewidth}[t]{
    >{\hsize=0.7\hsize}X
    >{\hsize=2.8\hsize}X 
    >{\hsize=1.0\hsize}X 
    >{\hsize=0.3\hsize}X 
    >{\hsize=0.2\hsize}X
  }
\toprule
\textbf{Data   Item} &
  \textbf{Short Description} &
  \textbf{Example} &
  \textbf{Value} &
  \textbf{RQs} \\ \midrule
Code &
  A unique identifier assigned to the paper &
  trabelsi2024\_magnet &
  Varied &
   \\
Title &
  The title of the research paper &
  Magnet: Method-Based Approach Using Graph Neural Network for   Microservices Identification &
  Varied &
   \\
Year &
  The year in which the   paper was published &
  2024 &
  Fixed &
   \\
Venue &
  The conference, journal, or workshop where the study was   published &
  International Conference on Software Architecture &
  Varied &
   \\
Authors &
  The individuals who contributed to conducting the study &
  Trabelsi, I and Moha, N and Guéhéneuc, YG and ... &
  Varied &
   \\
Migration Phase &
  The specific phase of system migration from monolith to   microservices &
  Identification &
  Fixed &
  RQ1 \\
Automated Task &
  Exact task automated during the migration phase &
  Identification &
  Varied &
  RQ1 \\
ML Integration &
  Specifies where machine learning (ML) is integrated &
  Clustering and semantic analysis &
  Varied &
  RQ1 \\
Input type &
  Defines the types of input used, such as source code, logs,   UML diagrams, etc. &
  Source code &
  Varied &
  RQ2 \\
Granularity of Data &
  The level of detail or precision in the data, such as files,   classes, lines of code, or logs &
  Methods &
  Fixed &
  RQ2 \\
Source of Data &
  Indicates the data source: open source, industrial projects,   user-generated, or collected &
  Open Source &
  Fixed &
  RQ2 \\
Data preprocessing &
  Steps to prepare data, including collection, cleaning,   normalization, and transformation &
  Static analysis with KDM and semantic analysis using word2vec &
  Varied &
  RQ2 \\
ML Technique &
  Indicates the machine learning techniques used &
  Deep Modularity Networks (DMoN) &
  Varied &
  RQ3 \\
Learning Approach &
  Indicates the learning approach: supervised, semi-supervised,   unsupervised, etc. &
  Unsupervised &
  Fixed &
  RQ3 \\
Feature Selection &
  Specifies the features used for model training &
  method calls, method bodies, and class structures &
  Varied &
  RQ3 \\
Evaluation Metrics &
  Specifies the metrics   used to evaluate the output &
  Precision, recall, f-measure, SMQ (Structural Modularity   Quality), CMQ (Conceptual Modularity Quality), CHM (Cohesion at Message level), and CHD (Cohesion at Domain level). &
  Varied &
  RQ4 \\
Benchmark &
  Indicates performance   benchmarks or comparisons with other systems &
  Compared against ServiceCutter and MicroMiner &
  Varied &
  RQ4 \\
Success Criteria &
  Specifies success   criteria, such as improved recall, quality, or performance &
  Improved modularity, functional independence, and reduced coupling. &
  Varied &
  RQ4 \\
Tool Availability &
  Indicates tool   availability: open-source, commercially available, or a proof of concept   (PoC) &
  Available, Open Source &
  Fixed &
  RQ4 \\
Challenges &
  Identifies the   challenges discussed in the paper &
  Generalizability   across different types of monoliths and data availability &
  Varied &
  RQ5 \\ \bottomrule
\end{tabularx}
}
\end{threeparttable}
\end{table*}

\section{Overview of the Selected Literature}

We analysed the primary studies (PSs) (presented in Table \ref{tab:overview_selected_literature} in appendix \ref{app:methodology}) to gain deeper insights into the selected set. In this section, we present our findings on publication trends, distribution of PSs by study type, databases used, and the publication venues associated with these PSs.
 \subsection{Publication Trends}
\begin{figure*}[ht]
 \includegraphics[width=\linewidth]{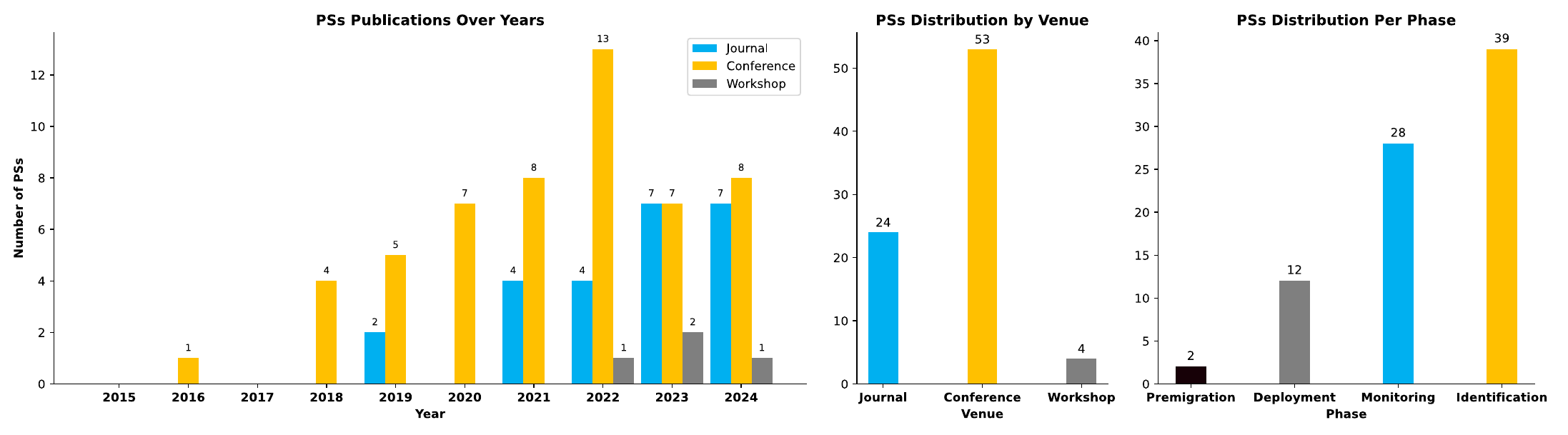}
 \caption{Distribution of PSs} 
 \label{fig:distributionOfPSs}
 \end{figure*}

 We analysed the publication trends of the selected primary studies (PSs). Notably, 17.2\% of the PSs were published in 2022, marking the year with the highest publication volume. An increasing trend in conference publications is observed from 2016 to 2022. In contrast, journal publications show a rise starting in 2021, with no journal publications recorded before this period except for two PSs in 2019. Only a few PSs were published in workshop proceedings, specifically in 2022 (1 PS), 2023 (2 PSs), and 2024 (1 PS). Figure \ref{fig:distributionOfPSs} shows the publication trends over the years for the PSs.
 \subsection{Distribution of PSs by Type} 
 We analysed the types of the selected primary studies (PSs). As shown in Figure \ref{fig:distributionOfPSs} most of the PSs are from conferences, comprising 65\% (53 PSs), followed by journals with 30\% (24 PSs). The smallest group is from workshops, accounting for only 5\% (4 PSs).
 The high conference publication rate may indicate rapid field growth, while journal articles add rigour. Limited workshop presence may reflect field maturity or reduced exploratory discussions. 
\subsection{Publication Venue}
We identified 72 distinct venues for the PSs. Most venues have a single PS, except for ASPLOS (2 PSs), ICWS (2 PSs), ISSRE (2 PSs), ASE (3 PSs), ICSA (3 PSs), and ICSOC (3 PSs). 
The broad distribution of PSs across different publication venues suggests that the research topic attracts interest from various subfields within software engineering and related areas. However, most studies remain monodisciplinary in nature, focusing predominantly on aspects of software architecture, migration engineering, and software modernisation.

\subsection{Databases for Selected PSs}
We observe that the majority of PSs were sourced from IEEE Xplore (37 PSs or 45.6\%), followed by Compendex (21 PSs or 25.9\%) and ACM Digital Library (10 PSs or 12.3\%). We observed the presence of PSs in other databases, such as Scopus (6 PSs or 7.4\%), ScienceDirect (3 PSs or 3.7\%), and Wiley (3 PSs or 3.7\%). The Web of Science database provided the least contribution with 1 PS (1.2\%).
This diversity in sources enhances the robustness of the literature review by ensuring a comprehensive collection of relevant studies.

For more visuals and tables, see Appendix~\ref{app:methodology}.

%% file: Sections/4-RQ1.tex
\section{Migration Phases Automated by ML (RQ1)}
\label{Section: rq1}

The migration of legacy systems to microservices involves a sequence of complex phases, many of which are now supported or automated by machine learning. In this section, we \revised{analyse} the phases of the migration process and identify the specific tasks that ML automates in this process. Following prior works on microservices migration: Abgaz et al.~\cite{abgaz2023decomposition}, Fritzsch et al.~\cite{fritzsch_springer_2019_monolith}, and Saucedo et al.~\cite{saucedo_elsevier_2024_migration}, we structured the migration process into five phases: Pre-migration, Identification, Packaging, Deployment, and Monitoring. The identification phase is the most frequently covered in our primary studies (PSs), with 48\% (n=39), followed by the monitoring phase at 35\% (n=28). The deployment phase is also notable, covered by 15\% of PSs (n=12), while the pre-migration phase is the least discussed, with only 2.24\% (n=2). Figure \ref{fig:distributionOfPSs} shows the number of PSs that target each phase.  We provide the definitions for all key terms related to RQ1, including migration phases and associated automated tasks, in Appendix \ref{app:definitions_phases} for clarity and consistency.




\subsection{Pre-migration}
The pre-migration phase focuses on evaluating the current approaches, methods, and tools used in the legacy system. This phase also involves planning the migration by identifying suitable strategies, defining objectives, and outlining the steps needed to ensure a smooth transition. Researchers have leveraged ML to automate tasks such as designing microservices, which assists in planning microservice architecture by analyzing legacy systems \citeP{nakazawa2018_visualization_tool}, and predicting success rates of migration, which uses predictive models to evaluate potential migration outcomes \citeP{alshammari2023_genetic_hpc_migration}.

\subsection{Identification}
The identification phase defines the boundaries of prospective microservices. This involves detecting functional modules, mapping dependencies, and clustering components with business requirements. ML techniques have been applied to automate tasks such as boundary identification, which detects functional boundaries within monolithic systems \citeP{gysel2016_service_cutter_decomposition}; clustering, which groups related components or services using ML-based clustering algorithms \citeP{kamimura2018_microservice_candidates}; and microservices identification, which automates the process of determining candidate microservices \citeP{zhong2023_spectral_clustering_industrial_legacy,daoud2021multi,saidi2022_structural_dependency_microservices,daoud2020_microservice_identification,faria2022_code_vectorization_microservices,sun2022_expert_system_identification,trabelsi2023_type_based_microservices,trabelsi2024_magnet_graph_nn,trabelsi2024_micromatic_automation}.

\subsection{Packaging}
The packaging phase encapsulates the identified components into functional microservices. This phase includes defining service interfaces, managing dependencies, and generating missing components. However, no studies in our review proposed the use of ML for this phase, highlighting a significant gap in the research landscape.

\subsection{Deployment}
The deployment phase focuses on deploying the created microservices into a target environment. This involves setting up orchestration tools, configuring infrastructure, and integrating with existing systems. ML techniques are increasingly used to automate tasks such as automated deployment, which automates the deployment of microservices into production environments \citeP{trabelsi2024_micromatic_automation,lv2022_deployment_edge_computing}; resource management, which ensures efficient use of infrastructure resources during deployment \citeP{khan2023_dynamic_resource_management,joseph2019_fuzzy_rl_allocation,luan2024_resource_optimization,gan2019_seer_performance_debugging}; resource allocation, which dynamically allocates resources to microservices based on workload demands \citeP{khan2023_dynamic_resource_management,shafi2024_cdascaler_autoscaling,song2023_chainsformer_latency_aware_provisioning,gan2022_practical_cloud_performance,luan2024_resource_optimization,yang2019_miras_resource_allocation_workflows,zeng2023_topology_adaptive_provisioning}; microservice autoscaling, which adjusts the scale of microservices to match performance requirements \citeP{tong2023_gma_autoscaling_edge_cloud}; microservice orchestration, which manages dependencies and interactions between microservices \citeP{li2022_score_resource_orchestration}; microservice placement, which decides optimal placements for microservices within the infrastructure \citeP{ray2023_microservice_placement_edge,joseph2019_fuzzy_rl_allocation}; and resource estimation, which estimates the resource needs of microservices to optimize deployment strategies \citeP{chow2022_deeprest_resource_estimation}.

\subsection{Monitoring}
The monitoring phase ensures the reliability and efficiency of the microservices-based system post-migration. This phase involves continuous performance tracking, anomaly detection, and dynamic resource management to meet changing workload demands. ML models enhance the monitoring process by enabling proactive detection of issues and optimising resource usage. Tasks automated in this phase include anomaly detection, which identifies abnormal behaviors or performance issues in microservices \citeP{shi2022_bsdg_anomaly_detection,zhang2022_deeptralog_combined_anomaly,chen2023_bert_htlg_detection}; performance analysis, which evaluates the performance metrics of deployed microservices \citeP{song2024_autonomous_fault_classification}; detection of failure types, which classifies failure types to facilitate troubleshooting \citeP{xu2023_heterogeneous_failure_diagnosis,sun2024_failure_localization_autoencoder}; fault diagnosis, which diagnoses root causes of failures to enable faster resolution \citeP{song2024_autonomous_fault_classification}; root cause analysis, which pinpoints the underlying issues causing anomalies or failures \citeP{gan2022_practical_cloud_performance,gan2021_sage_ml_performance_debugging}; privacy risks detection, which identifies potential privacy concerns in microservice interactions \citeP{chou2021_security_privacy_nn}; and sanity checks, which verifies the correctness of system states and responses \citeP{chow2022_deeprest_resource_estimation}.


%% file: Sections/5-RQ2.tex
\section{Inputs used by ML migration approaches(RQ2)}
\label{Section: rq2}

\begin{figure}[ht]
 \includegraphics[width=\linewidth]{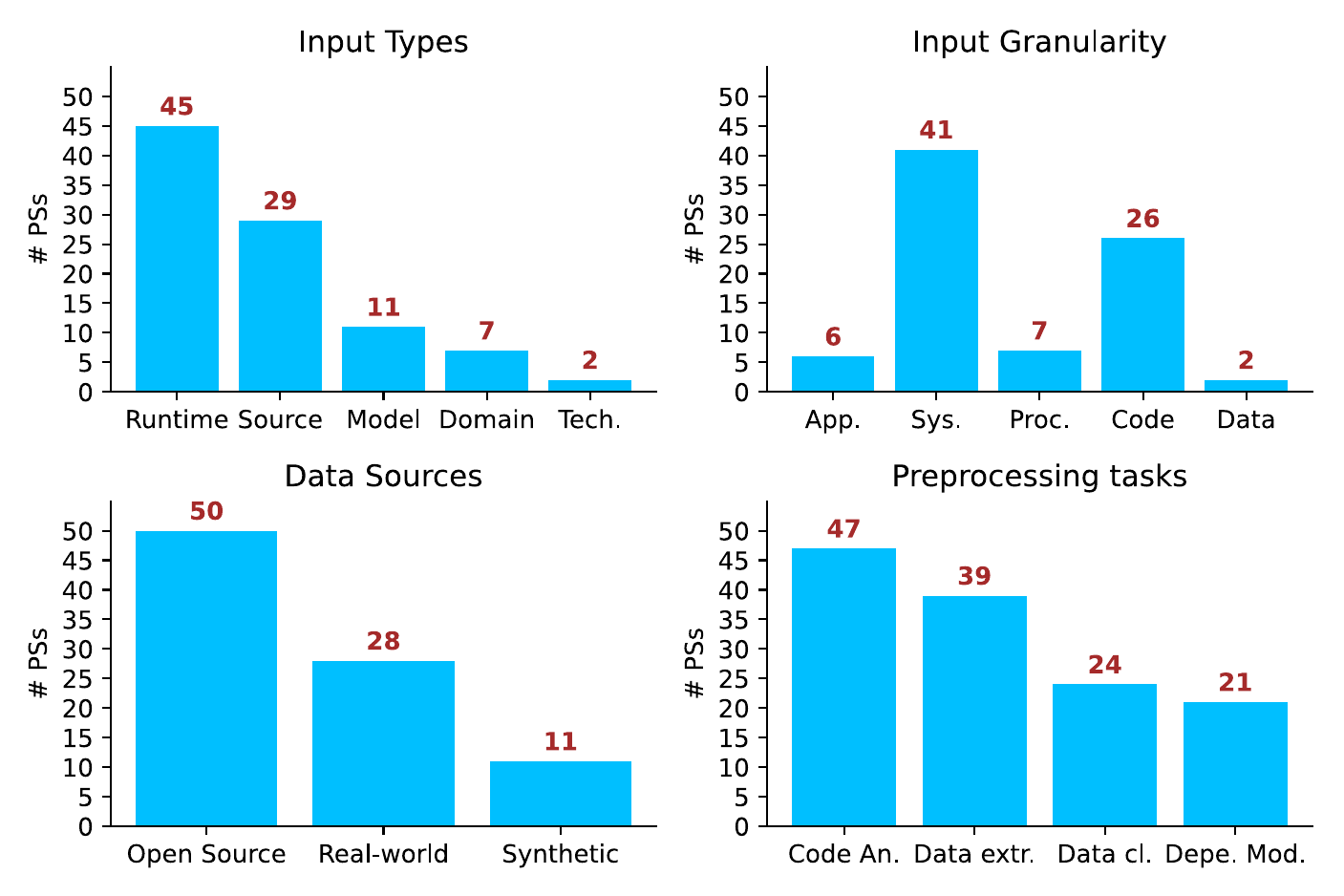}
 \caption{Inputs and Preprocessing Tasks in PSs}
 \label{fig:RQ2}
 \end{figure}

It is essential to understand the types of inputs collected and processed in the migration process to identify how machine learning techniques are applied. To this end, we categorized inputs along four dimensions: type, granularity, source, and preprocessing method. These categories were not predefined, but emerged inductively from our analysis of the primary studies. Figure \ref{fig:RQ2} summarizes data we extracted in our PSs.

The resulting taxonomy allowed us to systematically examine the role and characteristics of inputs used in ML-based migration approaches. We provide the definitions for all key terms, including input types, granularity levels, data sources, and preprocessing categories, in Appendix~\ref{app:definitions_rq2} for clarity and consistency.

\subsection{Input Types and Their Role in Migration}
In this section, we classify the input types identified in the primary studies (PSs) into five distinct groups which illustrates their use in the migration phases. Our analysis reveals that runtime artifacts are the most commonly used inputs, appearing in 55\% of the PSs (\(n=45\)), emphasizing their critical role in understanding system performance during migration. Source Artifacts, such as source code and configuration files, are the second most frequently used, accounting for 35\% of PSs (\(n=29\)). In contrast, domain artifacts (\(n=7\)) and model artifacts (\(n=11\)), provide valuable high-level system descriptions, while technical artifacts are the least common, found in only 2\% of PSs (\(n=2\)).

\subsubsection{Domain Artifacts}
Domain artifacts capture high-level information about the system's business and functional requirements, aiding in the understanding and identification of microservices. These include API documentation, which represents details about the available APIs and their interactions. For example, \citeP{sun2022_expert_system_identification} highlighted the use of the OpenAPI specification of the legacy system's RESTful APIs to identify microservices. Quality of Service (QoS) constraints represent parameters that must be maintained during migration. The authors of \citeP{chou2021_security_privacy_nn} used service communication traces to meet predefined QoS requirements, such as availability and reliability. Architecture recommendations, as provided by \citeP{selmadji2020_transition_microservices}, offer expert suggestions for migration strategies and component separation. Functional descriptions, such as those provided by \citeP{morais2021_ontology_microservices}, represent high-level descriptions of system functionalities, capturing technological attributes and relationships between services.

\subsubsection{Runtime Artifacts}
Runtime artifacts are derived from the system's runtime behavior and are crucial for understanding performance and operational characteristics. These include resource metrics, which focus on information about resource consumption, such as memory and CPU usage. The study by \citeP{chow2022_deeprest_resource_estimation} explored the use of API traffic logs to analyze the runtime behavior of systems. Similarly, \citeP{xu2023_heterogeneous_failure_diagnosis} leveraged system metric data, such as CPU, memory, and disk usage, along with failure propagation information, to identify resource bottlenecks. \citeP{du2018_anomaly_container_microservices} used memory utilisation, network latency, and packet loss to detect anomalies in containerised microservices, while \citeP{shafi2024_cdascaler_autoscaling} used CPU usage metrics to analyse performance across different stages of system operation. Monitoring metrics are another key runtime artifact. \citeP{kong2024_fault_localization_span} introduced SpanGraph, a tool that leverages monitoring metrics, trace logs, and configuration files to construct directed graphs representing microservice interactions. Performance metrics, such as response time, throughput, and latency, are also critical. \citeP{santos2024_performance_forecast_microservices} used time series data of performance metrics to predict system performance. Trace logs provide detailed insights into the execution flow of the system, as demonstrated by \citeP{chen2022_tracegra_anomaly_detection}, who used trace logs and performance metrics for anomaly detection. Workloads, which describe system usage scenarios and patterns, are used by \citeP{abdullah2019_autoscaling_policies} to predict response time of microservices.

\subsubsection{Model Artifacts}
Model artifacts describe the system from an architectural or design perspective, often providing an abstract representation. Business process models (BPMN) are used to capture the workflows supported by the system and align microservices with organisational functions. Several studies (e.g., \citeP{daoud2020_microservice_identification_business, daoud2020_microservice_identification, saidi2022_structural_dependency_microservices, daoud2021multi}) leverage BPMN to support microservice identification, demonstrating that business process understanding can guide the decomposition of monolithic systems into well-bounded services.
UML diagrams, which represent details of the system design, are also commonly used. For instance, \citeP{zhang2022_deeptralog_combined_anomaly} combined code, logs, and UML diagrams to provide unique insights for anomaly analysis, leveraging the visual representation of system design to identify issues. User stories are another key model artifacts, used by \citeP{vera2021_microservices_backlog_genetic} to guide the organisation of microservice development tasks.

\subsubsection{Source Artifacts}
Source artifacts pertain to the actual software and its configuration, offering insights into the system's implementation. For example, \citeP{trabelsi2024_micromatic_automation} used source code to identify microservices, while \citeP{nakazawa2018_visualization_tool} analyzed source code files from a monolithic application to generate a calling-context tree for better system understanding and transformation. Configuration files are used by \citeP{kong2024_fault_localization_span} alongside trace logs and monitoring metrics to construct SpanGraph, a directed graph representing microservice interactions for fault localization. Data files are used by \citeP{song2023_chainsformer_latency_aware_provisioning} to support latency-aware provisioning.

\subsubsection{Technical Artifacts}
They provide the supporting information necessary for system operation. These include server information, which details the infrastructure supporting the system. For example, \citeP{ray2023_microservice_placement_edge} used service radius and capacity, as well as runtime traces to predict microservice placement.

\subsection{Input Granularity}
The granularity of input data, ranging from high-level abstractions like system architecture diagrams to fine-grained elements such as lines of code or individual function calls, plays a critical role in determining the precision and scope of machine learning tasks. We organize the granularity levels in descending order of abstraction: application-level, system-level, process-level, code-level, and data-level. Application-level granularity (7\%, n=6) examines higher-level abstractions, with \citeP{bajaj2024_gtmicro_nlp_microservices} using use case-level data for workflow modeling and \citeP{zhang2022_deeptralog_combined_anomaly} analyzing application files for anomaly detection. URI and API-level granularity, as in \citeP{abdullah2019_unsupervised_web_decomposition} and \citeP{al_debagy2019_decomposition_method}, support system decomposition and service identification. System-level granularity is the most frequently used (51\%, n=41), as seen in \citeP{lv2022_deployment_edge_computing} and \citeP{lv2024_graph_rl_deployment}, which monitor and analyze system components and interactions. Process-level granularity (9\%, n=7) examines workflows, with \citeP{al_debagy2019_decomposition_method} using API operation-level data for decomposition and \citeP{daoud2020_microservice_identification} using activity-level granularity for workflow modeling. Code-level granularity (32\%, $n=26$) targets classes and methods for tasks like refactoring and microservice identification, as seen in \citeP{rathod2023_industry4_refactoring} and \citeP{trabelsi2024_magnet_graph_nn}. Finally, data-level granularity (2\%, $n=2$) is the least explored and focuses on stored information, such as table-level data in \citeP{romani2022_data_centric_identification} and entity-level representations in \citeP{gysel2016_service_cutter_decomposition}.

\subsection{Data Sources}
Our analysis shows that open-source datasets are the most widely used, representing 62\% of PSs (n=50), showcasing the reliance on publicly available repositories for experimentation and validation. For example, \citeP{cao2022_domain_oriented_decomposition} validated their approach on on open-source applications like JPetStore. Real-world data, which provides insights from operational systems, is used in 35\% of PSs (n=28). For instance, \citeP{li2021_deepstitch_cross_layer} used the AIOps Challenge dataset to detect anomalies and faults in microservice systems. Meanwhile, synthetic data, generated to simulate real-world conditions, is utilized in 14\% of PSs (n=11). For example, \citeP{gan2021_sage_ml_performance_debugging} combined synthetic data from Apache Thrift with open-source datasets like DeathStarBench to debug performance in microservice environments, while \citeP{alshammari2023_genetic_hpc_migration} collected data using a survey and expert interviews to optimize migration strategies.

\subsection{Preprocessing Tasks}

Preprocessing ensures the quality and usability of input data for machine learning models. Code analysis is the most frequently employed technique (58\%, \(n=47\)), underscoring the importance of understanding system structure and semantics. Typical approaches rely on static analysis tools such as JavaParser, which parses source code to extract classes, methods, and dependencies, providing structured inputs for machine learning models. Data extraction, used in 48\% of PSs (\(n=39\)), captures essential operational and dependency information, such as parsing execution traces to build graph representations, as in \citeP{chen2023_dynamic_static_features}. Data cleaning, employed in 30\% of PSs (\(n=24\)), improves data quality through noise reduction and normalization. For example, \citeP{chen2022_tracegra_anomaly_detection} applied density-based clustering to filter noisy data points, while \citeP{chow2022_deeprest_resource_estimation} normalised trace data for compatibility with further analysis. Dependency modeling, found in 26\% of PSs (\(n=21\)), structures relationships for better analysis. For instance, \citeP{liu2020_dual_clustering_microservices} used vectorisation and embedding for clustering, and \citeP{sooksatra2022_vae_gnn_approach} built dependency graphs using normalised matrices for analysis with Variational Autoencoders (VAE). Modeling and task structuring, as in \citeP{lv2024_graph_rl_deployment}, represented microservice dependencies using adjacency matrices, while \citeP{luan2024_resource_optimization} captured features like CPU usage and request counts for workload analysis.

 


%% file: Sections/6-RQ3.tex
\section{ML approaches applied by researchers (RQ3)}
\label{Section: rq3}
\begin{figure}[ht]
 \includegraphics[width=\linewidth]{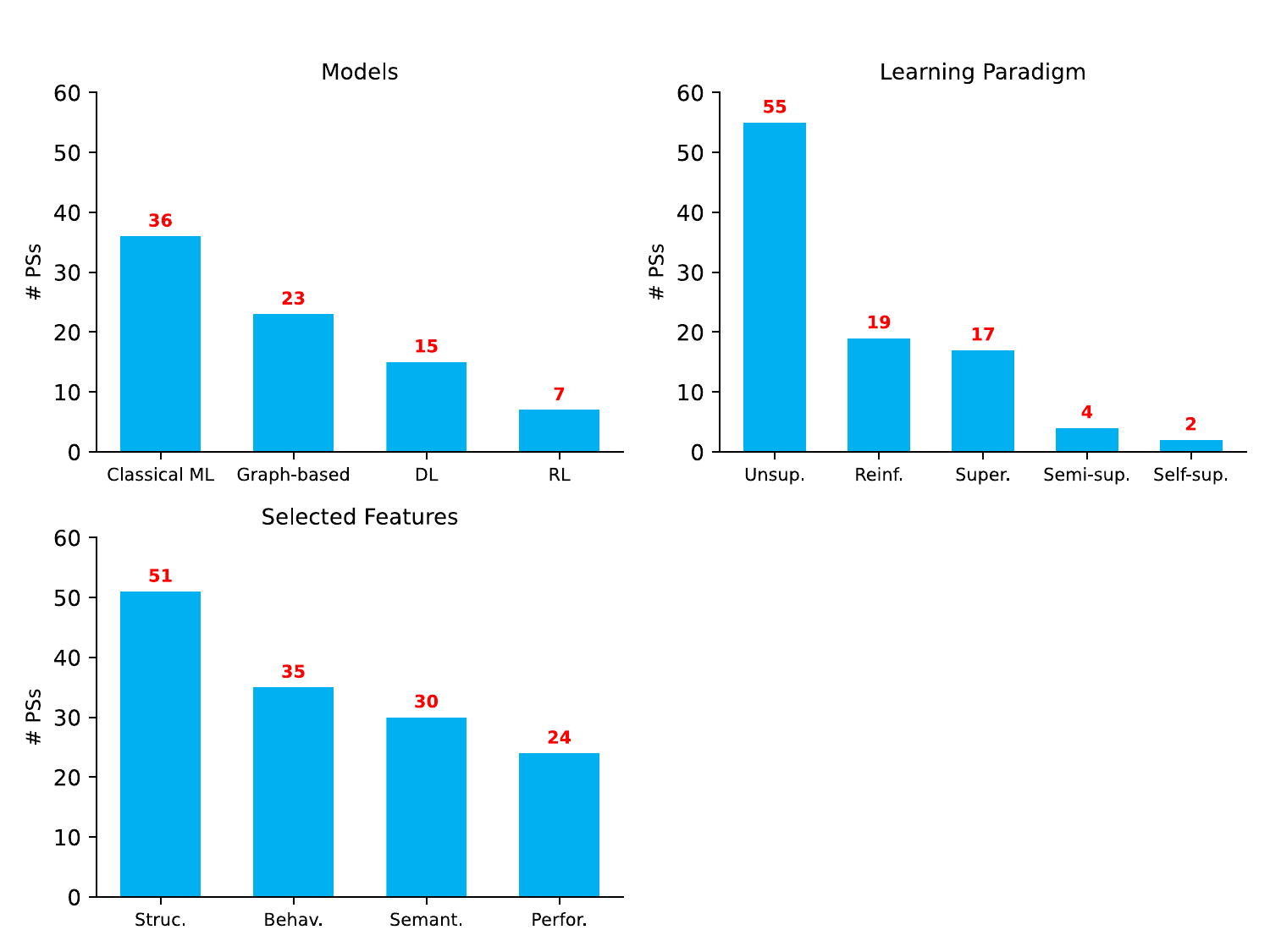}
 \caption{ ML Models, Learning Paradigms, and Features in PSs}
 \label{fig:RQ3}
 \end{figure}

In this section, we discuss the key ML models used in microservices migration, their learning paradigms, and the features that enhance model performance.
 Figure \ref{fig:RQ3} \revised{summarises the data we extracted from our PSs.}

We provide the definitions for all key terms related to RQ3 in Appendix \ref{app:definitions_rq3} for clarity and consistency.

\subsection{Models}
Different machine learning models have been applied to address various phases of the microservices migration process. Based on our systematic analysis of the literature and aligned with established machine learning taxonomies \cite{murphy2022probabilistic, wu2020comprehensive}, we categorise these models into four distinct classes: Classical machine learning, Deep learning, Graph-based models and Reinforcement learning. Classical ML models were the most frequently used (44.40\%, \(n=36\)), followed by graph-based models (28.40\%, \(n=23\)), deep learning (18.50\%, \(n=15\)), and reinforcement learning (8.6\%, \(n=7\)).

\subsubsection{Classical Machine Learning}  
Classical machine learning techniques are predominantly applied to structured datasets. Nevertheless, several models, such as Naive Bayes or Support Vector Machines, also perform well on unstructured data (e.g., text) once transformed into feature vectors through methods like TF-IDF or word embeddings. These techniques are widely applied to automate tasks such as service identification, dependency analysis, and performance prediction. Classification techniques, such as Support Vector Machines \citeP{trabelsi2023_type_based_microservices}, Random Forest \citeP{daoud2021multi}, and Naive Bayes \citeP{al2021microservice}, are used to categorise components and analyse dependencies. Regression techniques, including Linear Regression \citeP{al2021microservice} and Support Vector Regression \citeP{zhong2023_spectral_clustering_industrial_legacy}, predict performance metrics and resource usage. Clustering methods, such as K-means \citeP{al2021microservice} and Density-Based Spatial Clustering of Applications with Noise \citeP{sellami2022_hierarchical_dbscan_microservices}, group components for service decomposition. Search-based techniques, like Genetic Algorithms \citeP{alshammari2023_genetic_hpc_migration}, optimise migration strategies.
\subsubsection{Graph-Based Models}  
Graph-based methods model component relationships as graphs, enabling service interaction and dependency analysis. Graph Convolutional Networks \citeP{mathai2021_hgnn_microservices_representation} and Graph Attention Networks \citeP{trabelsi2024_magnet_graph_nn} aggregate features from neighbouring nodes to improve graph representations. Variational Graph Autoencoders \citeP{sooksatra2022_vae_gnn_approach} combine graph convolutional networks with variational inference for tasks like link prediction.
\subsubsection{Deep Learning}  
Deep learning techniques leverage neural networks to analyse complex data structures. Autoencoders and Variational Autoencoders \citeP{sooksatra2022_vae_gnn_approach} are used for dimensionality reduction and feature learning. Recurrent networks, such as Long Short-Term Memory \citeP{li2021_deepstitch_cross_layer} and Gated Recurrent Units \citeP{zhang2022_deeptralog_combined_anomaly}, process sequential data for anomaly detection. Transformers, including Bidirectional Encoder Representations from Transformers \citeP{chen2023_bert_htlg_detection} and CodeBERT \citeP{faria2022_code_vectorization_microservices}, excel in natural language and code analysis tasks.

\subsubsection{Reinforcement Learning}  
Reinforcement learning learns optimal strategies through interactions with the environment. Fuzzy Q-Learning \citeP{joseph2019_fuzzy_rl_allocation} used to handles uncertainty in resource allocation, while Deep Q-Learning \citeP{abdullah2019_autoscaling_policies} and Deep Deterministic Policy Gradient \citeP{lv2024_graph_rl_deployment} used to optimise autoscaling and resource usage. Multi-Agent Deep Deterministic Policy Gradient \citeP{ray2023_microservice_placement_edge} improves service placement in edge computing environments.
\subsection{Learning paradigm}
The learning paradigms applied in the studies include unsupervised learning, which is the most popular (67.9\%, \(n=55\)), used for tasks like microservice extraction, clustering, and anomaly detection \citeP{trabelsi2023_type_based_microservices, shahini2024_autoencoder_anomaly_detection}; reinforcement learning (23.46\%, \(n=19\)), primarily for resource provisioning and deployment optimisation \citeP{song2023_chainsformer_latency_aware_provisioning, lv2024_graph_rl_deployment}; supervised learning (20.99\%, \(n=17\)), applied in anomaly detection and resource optimisation \citeP{tan2024_maad_anomaly_detection, abdullah2019_autoscaling_policies}; semi-supervised learning (4.94\%, \(n=4\)), which combines labelled and unlabeled data to refine service boundaries \citeP{al2021microservice}; and self-supervised learning (2.47\%, \(n=2\)), which generates pseudo-labels for tasks like failure localisation and resource provisioning \citeP{sun2024_failure_localization_autoencoder, zeng2023_topology_adaptive_provisioning}.

\subsection{Selected Features}  
Machine learning models in these PSs rely on diverse feature types to analyze, predict, and optimize different aspects of the migration process. Structural features are the most popular (62.96\%, \(n=51\)), including class dependencies \citeP{nitin2022_cargo_dependency_analysis}, method calls \citeP{kamimura2018_microservice_candidates}, data dependencies \citeP{romani2022_data_centric_identification}, transactional dependencies \citeP{daoud2021multi}, and call graph dependencies \citeP{qian2023_graph_clustering_extraction}. Behavioral features (43.21\%, \(n=35\)) capture runtime interactions, such as invocation paths \citeP{kamimura2018_microservice_candidates}, log events \citeP{shahini2024_autoencoder_anomaly_detection}, contextual log entries \citeP{shahini2024_autoencoder_anomaly_detection}, and temporal patterns \citeP{shahini2024_autoencoder_anomaly_detection}. Semantic features (37.04\%, \(n=30\)) focus on the meaning of system components, leveraging semantic embeddings \citeP{trabelsi2023_type_based_microservices}, function names \citeP{saidi2023_ddd_migration_microservices}, and API descriptions \citeP{kamimura2018_microservice_candidates}. Performance features (29.63\%, \(n=24\)) measure system efficiency, including CPU usage \citeP{bajaj2024_gtmicro_nlp_microservices}, memory usage \citeP{sooksatra2022_vae_gnn_approach}, network traffic \citeP{liu2024_migration_graph_nn}, response times \citeP{bajaj2024_gtmicro_nlp_microservices}, and availability \citeP{nunes2019_transactional_contexts}. These features collectively enable tasks such as microservice identification, dependency analysis, anomaly detection, and performance optimisation.

%% file: Sections/7-RQ4.tex
\section{evaluation of ML approaches (RQ4)}
\label{Section: rq4}

\begin{figure}[ht]
 \includegraphics[width=\linewidth]{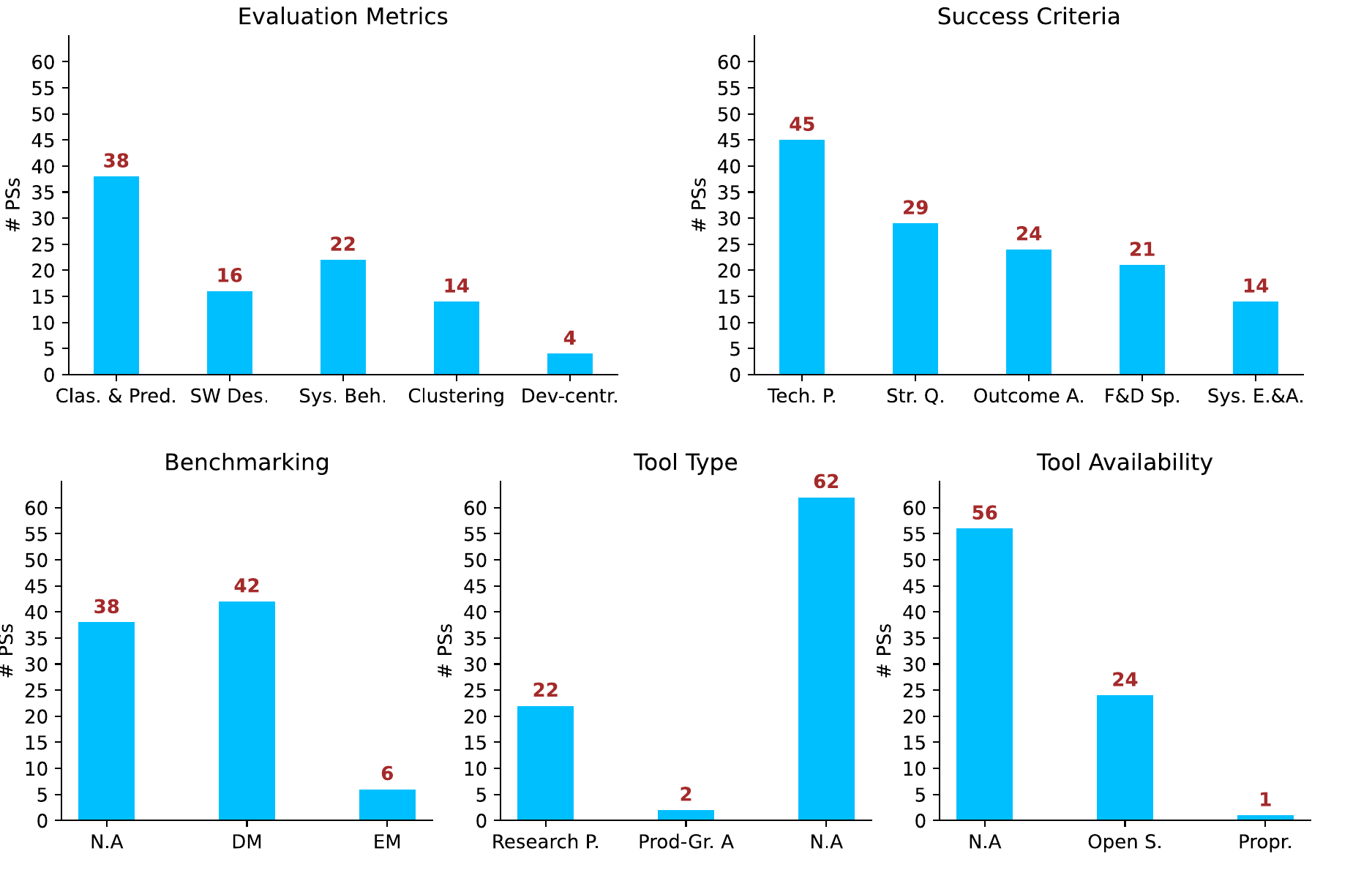}
 \caption{ Evaluation of ML Approaches in PSs} 
 \label{fig:RQ4}
 \end{figure}

In this section, we begin by discussing the various evaluation metrics commonly employed to assess the performance of machine learning models in migration scenarios. Next, we explore the conducted comparative analyses. Following this, we outline the success criteria adopted in the literature. Finally, we will discuss the tools availability of those works.
Although the \textit{Data Sources} category is discussed in Section~\ref{Section: rq2} as part of input characterization, it also plays an important role in RQ4. The same data are frequently used not only as inputs to the ML-based migration techniques but also as a foundation for their evaluation and validation. 
 Figure \ref{fig:RQ4} summarizes data we extracted in our PSs.

We provide the definitions for all key terms related to RQ4 in Appendix \ref{app:definitions_rq4}.

\subsection{Evaluation Metrics}
Evaluation metrics are categorized into five key areas: classification and prediction metrics (46\%, n=38), software design metrics (19\%, n=16), system behavior metrics (27\%, n=22), clustering metrics (17\%, n=14), and developer-centric metrics (5\%, n=4). Each category addresses specific aspects of microservices migration.

\subsubsection{Classification and Prediction Metrics}
These metrics assess model performance in labeling and prediction tasks, primarily during the identification phase. Precision measures the accuracy of positive predictions, with variations like Precision@K (PR@K) and Mean Average Precision (MAP) used for ranked predictions or multiple queries. It is commonly applied in monitoring for anomaly detection, as seen in \citeP{du2018_anomaly_container_microservices, shahini2024_autoencoder_anomaly_detection, chen2023_bert_htlg_detection, shi2022_bsdg_anomaly_detection}. Recall evaluates the model's ability to identify all relevant instances, particularly in anomaly detection and fault identification tasks, as demonstrated in \citeP{du2018_anomaly_container_microservices, al2021microservice}. The F1-Score balances precision and recall, making it critical for monitoring and service identification tasks, as shown in \citeP{chen2023_dynamic_static_features, du2018_anomaly_container_microservices, bajaj2020_partial_migration_cloud_native, shahini2024_autoencoder_anomaly_detection, daoud2021multi}. Accuracy represents the overall correctness of predictions, with variations like Top-K Accuracy used in multi-class classification tasks, as seen in \citeP{zhong2023_spectral_clustering_industrial_legacy, tan2024_maad_anomaly_detection, nakazawa2018_visualization_tool, tong2023_gma_autoscaling_edge_cloud}. The Area Under the ROC Curve (AUC) evaluates binary classification models, particularly in anomaly detection, as highlighted in \citeP{daoud2020_microservice_identification_business, kalia2020_mono2micro}. Finally, the Matthews Correlation Coefficient (MCC) provides a balanced measure of classification quality, especially in imbalanced datasets, as used in \citeP{shahini2024_autoencoder_anomaly_detection}.

\subsubsection{Clustering Metrics}
These metrics evaluate the quality of service clustering during the identification phase. The Dunn Index measures the separation between clusters, as demonstrated in \citeP{zhong2023_spectral_clustering_industrial_legacy, sellami2022_hierarchical_dbscan_microservices, bajaj2020_partial_migration_cloud_native, chen2023_dynamic_static_features}. The Silhouette Score assesses the similarity of objects within their cluster compared to others, as seen in \citeP{sellami2022_hierarchical_dbscan_microservices, tong2023_gma_autoscaling_edge_cloud, daoud2020_microservice_identification_business, chou2021_security_privacy_nn}. Newman-Girvan Modularity evaluates the strength of network division into clusters, as highlighted in \citeP{saidi2023_ddd_migration_microservices, zhong2023_spectral_clustering_industrial_legacy, chen2023_dynamic_static_features, tan2024_maad_anomaly_detection}. Non-Extreme Distribution ensures balanced cluster sizes \citeP{mathai2021_hgnn_microservices_representation, qian2023_graph_clustering_extraction, rathod2023_industry4_refactoring, sooksatra2022_vae_gnn_approach}. Maximum Cluster Size limits the largest cluster size to avoid granularity issues \citeP{nunes2019_transactional_contexts}. The Number of Singleton Clusters counts clusters with a single element to prevent excessive fragmentation \citeP{nunes2019_transactional_contexts}.

\subsubsection{System Behavior Metrics}
These metrics assess operational performance, resource utilization, and scalability during and after migration. Response Time measures the time taken for the system to respond to a request, as highlighted in \citeP{al2021microservice, tan2024_maad_anomaly_detection, bajaj2020_partial_migration_cloud_native, du2018_anomaly_container_microservices}. Resource Utilization evaluates the effectiveness of resource use, as seen in \citeP{sellami2022_hierarchical_dbscan_microservices, tong2023_gma_autoscaling_edge_cloud, saidi2023_ddd_migration_microservices, chou2021_security_privacy_nn}. Energy Consumption assesses the total energy required for operations, as demonstrated in \citeP{daoud2021multi, tan2024_maad_anomaly_detection, bajaj2020_partial_migration_cloud_native, nakazawa2018_visualization_tool}. The SLA Violation Rate measures the percentage of time a system fails to meet its Service Level Agreement obligations, as shown in \citeP{saidi2023_ddd_migration_microservices, sellami2022_hierarchical_dbscan_microservices, daoud2020_microservice_identification_business, tong2023_gma_autoscaling_edge_cloud}. Scalability evaluates the system's capacity to maintain or improve performance under increased workload, as in \citeP{daoud2021multi, zhong2023_spectral_clustering_industrial_legacy, chen2023_dynamic_static_features, chou2021_security_privacy_nn}.

\subsubsection{Software Design Metrics}
These metrics evaluate software quality during the evaluation phase, focusing on maintainability and modularity. Cohesion measures the relatedness of functionalities within a service, as demonstrated in \citeP{daoud2021multi, sellami2022_hierarchical_dbscan_microservices, bajaj2020_partial_migration_cloud_native, chou2021_security_privacy_nn}. Coupling assesses dependencies between services, as seen in \citeP{al2021microservice, zhong2023_spectral_clustering_industrial_legacy, saidi2023_ddd_migration_microservices, tan2024_maad_anomaly_detection}. The Granularity Metric evaluates service size and scope, as highlighted in \citeP{sellami2022_hierarchical_dbscan_microservices, tong2023_gma_autoscaling_edge_cloud, chen2023_dynamic_static_features, nakazawa2018_visualization_tool}. Structural Modularity measures the degree of system decomposition into independent components, as shown in \citeP{zhong2023_spectral_clustering_industrial_legacy, daoud2020_microservice_identification_business, bajaj2020_partial_migration_cloud_native, du2018_anomaly_container_microservices}. Cognitive Complexity assesses code understandability for developers, as demonstrated in \citeP{saidi2023_ddd_migration_microservices, sellami2022_hierarchical_dbscan_microservices, nakazawa2018_visualization_tool, tan2024_maad_anomaly_detection}.

\subsubsection{Developer-Centric Metrics}
These metrics evaluate alignment with business requirements and developer feedback during the validation phase. Developer Validation assesses qualitative insights from developers, as seen in \citeP{daoud2021multi, sellami2022_hierarchical_dbscan_microservices, bajaj2020_partial_migration_cloud_native, nakazawa2018_visualization_tool}. Closeness to Manual Expert Analysis compares automated classifications to expert evaluations, as highlighted in \citeP{sellami2022_hierarchical_dbscan_microservices, daoud2020_microservice_identification_business, chen2023_dynamic_static_features, tong2023_gma_autoscaling_edge_cloud}.
\subsection{Benchmarking}
The reviewed PSs demonstrate a strong emphasis on benchmarking to evaluate and compare the effectiveness of their approaches. Direct method comparisons are the most commonly conducted, appearing in 51.85\% of the PSs (n=42), while evolutionary method comparisons are less frequent, accounting for only 7.41\% of the PSs (n=6).

\subsubsection{Direct Method Comparisons}
This category includes methodologies that assess the performance of ML models against existing approaches and techniques. Most PSs utilized comparisons with the State of the Art to benchmark current techniques against the latest advancements in the field, as well as against traditional rule-based and heuristic methods. For example, \citeP{abdullah2019_autoscaling_policies, liu2024_migration_graph_nn, al2021microservice, bajaj2024_gtmicro_nlp_microservices, chen2023_dynamic_static_features, trabelsi2023_type_based_microservices, trabelsi2024_magnet_graph_nn} highlight comparisons across ML-based approaches and traditional techniques. Additionally, some PSs focused on the effectiveness of widely used algorithms and techniques, such as clustering algorithms like K-Means, DBSCAN, and their variations. These were explored in studies like \citeP{cai2021_tracemodel_microservices, chen2022_edge_attention_localization, dehghani2022_migration_model_driven_rl}.

\subsubsection{Evolutionary Method Comparisons}
This category emphasizes the development of new methodologies built upon previous research. A few PSs aimed at enhancing existing Research by improving established techniques through the integration of novel features or adaptations for specific microservices migration scenarios. For instance, \citeP{daoud2020_microservice_identification, mathai2021_hgnn_microservices_representation} demonstrated enhancements to clustering and service identification methodologies. Other PSs explored combinatory approaches, combining different methodologies to leverage their strengths and improve migration outcomes. Examples include hybrid methods that integrate clustering with deep learning models for better service decomposition, as seen in \citeP{chen2023_dynamic_static_features, chow2022_deeprest_resource_estimation}.
\subsection{Success Criteria}
The success of ML-based migration approaches is determined by various criteria. Technical performance is the most emphasized, appearing in 55.56\% of the PSs (n=45), with key indicators including improvements in precision, recall, and resource optimization. Structural quality is considered in 35.80\% of the PSs (n=29), focusing on modularity and cohesion. Outcome alignment with standards and baselines is assessed in 29.63\% of the PSs (n=24). Functional and domain-specific success is addressed in 25.93\% of the PSs (n=21), while system effectiveness and adaptability are evaluated in only 17.28\% of the PSs (n=14).

\subsubsection{Technical Performance Metrics}
These metrics focus on the operational effectiveness of the ML models and their impact on system performance. Key indicators include increased precision, recall, and F1-Score, which demonstrate the effectiveness of the proposed ML-based approach in performing specific tasks, as shown in \citeP{al_debagy2019_decomposition_method, bajaj2024_gtmicro_nlp_microservices, song2024_autonomous_fault_classification, shi2022_bsdg_anomaly_detection}. Reduced response times and latency indicate improved system performance, leading to better user experiences and satisfaction, as highlighted in \citeP{li2022_score_resource_orchestration, lv2022_deployment_edge_computing, nitin2022_cargo_dependency_analysis}. Enhanced efficiency, reflecting the ability of the migration process to operate with limited resources while achieving desired outcomes, is another critical indicator, as seen in \citeP{cao2022_domain_oriented_decomposition, joseph2019_fuzzy_rl_allocation}. Increased throughput, which measures the system’s capacity to handle requests, reflects the effectiveness of the ML approach in optimizing service delivery, as demonstrated in \citeP{zhong2023_spectral_clustering_industrial_legacy, li2021_deepstitch_cross_layer}.

\subsubsection{Outcome Alignment with Standards and Baselines}
This criterion evaluates how well the outcomes of the ML-driven migration align with established standards and benchmarks. Achieving a defined baseline of performance metrics is crucial for evaluating the success of the migration, as improvements over this baseline indicate the effective use of ML methodologies. For example, \citeP{sun2022_expert_system_identification, kamimura2018_microservice_candidates, gysel2016_service_cutter_decomposition} demonstrated surpassing baseline metrics as a success criterion. Additionally, developer architectural alignment ensures that the architecture of the migrated services aligns with developer expectations, with positive feedback from developers serving as a key indicator of success, as seen in \citeP{nunes2019_transactional_contexts, morais2021_ontology_microservices}.

\subsubsection{Structural Quality}
This category focuses on the integrity and quality of the microservices architecture. Increased modularity ensures that microservices are independently deployable and maintainable, allowing for easier updates and scaling, as noted by \citeP{daoud2021multi, desai2021_gnn_outlier_refactoring}. Increased cohesion assesses the relatedness of functionalities within a service, as highlighted in \citeP{mathai2021_hgnn_microservices_representation, vera2021_microservices_backlog_genetic}. Decreased coupling measures the reduction in dependencies between services \citeP{cao2022_domain_oriented_decomposition, trabelsi2024_magnet_graph_nn}.

\subsubsection{System Effectiveness and Adaptability}
This criterion evaluates how well the system performs and its ability to adapt to changing requirements. Increased scalability measures the system's ability to scale resources up or down as needed to accommodate varying workloads, as emphasized in \citeP{lv2024_graph_rl_deployment, santos2024_performance_forecast_microservices}. Increased resource utilization aims to maximize the effectiveness of resource use during migration, as shown in \citeP{zhang2022_deeptralog_combined_anomaly, tan2024_maad_anomaly_detection}. Increased automation evaluates the extent to which processes are automated, reducing manual intervention and enhancing efficiency, as highlighted in \citeP{trabelsi2024_micromatic_automation}.

\subsubsection{Functional and Domain-Specific Success}
This criterion evaluates how well ML models meet domain-specific requirements and objectives. Increased anomaly detection quality enhances the reliability of the migration process and operational stability, as demonstrated in \citeP{shahini2024_autoencoder_anomaly_detection, chen2023_bert_htlg_detection}. Reduced QoS violations indicate that the migrated services are performing within acceptable parameters, as seen in \citeP{gan2022_practical_cloud_performance, abdullah2019_autoscaling_policies}. Increased cost efficiency reflects the ability to achieve desired outcomes with minimal financial expenditure, as highlighted in \citeP{shafi2024_cdascaler_autoscaling, abdullah2019_unsupervised_web_decomposition}.

\subsection{Tool Types and Availability}
\label{sub:tool_availabilty}
Research prototypes are the most commonly reported tool type, appearing in 27.18\% of the PSs (n=22), while only 2.47\% (n=2) describe production-grade applications. Regarding availability, 69.14\% of the PSs (n=56) did not provide information on tool availability, whereas 30\% (n=24) described open-source tools. Proprietary tools are mentioned in one PS.

\subsubsection{Tool Types}
Research prototypes, developed within research projects for validation purposes, are often partial implementations (20 PSs). Examples include tools proposed in \citeP{al_debagy2021_dependencies_based_decomposition, bajaj2020_partial_migration_cloud_native, desai2021_gnn_outlier_refactoring, gan2019_seer_performance_debugging, nitin2022_cargo_dependency_analysis}. In contrast, production-grade applications are fully developed, robust, and scalable tools ready for real-world deployment, as highlighted in \citeP{gysel2016_service_cutter_decomposition}.

\subsubsection{Tool Availability}
Open-source tools, made publicly available through platforms like GitHub, enable reuse and collaboration, as seen in \citeP{cao2022_domain_oriented_decomposition, desai2021_gnn_outlier_refactoring, nitin2022_cargo_dependency_analysis, trabelsi2024_magnet_graph_nn}. Proprietary tools, such as those discussed in \citeP{chow2022_deeprest_resource_estimation}, have restricted access and are not publicly available. However, the majority of PSs (74\%) provide no information on tool availability, limiting reproducibility and reuse.

%% file: Sections/8-RQ5.tex
\section{Challenges in using ML approaches (RQ5)}
\label{Section: rq5}

Machine learning has shown potential in supporting various phases of microservices migration. However, several challenges limit its effectiveness and broader adoption. These challenges were derived through manual analysis of the primary studies (PSs), specifically by reviewing:
i) the Discussion section,
ii) the Threats to Validity section,
iii) the Future Work section, and
iv) the Conclusion section of each study. Only challenges explicitly reported by the study authors, rather than inferred, were considered to ensure objectivity. The extracted challenges range from technical complexities to data quality issues, scalability concerns, integration difficulties, and specialized skill requirements. This section discusses these challenges and explores possible solutions.

\subsection{Technical Complexities}
One of the primary challenges in applying ML to software migration is handling the variability and complexity of legacy systems. Identifying service boundaries within monolithic architectures remains a difficult task, often leading to suboptimal service decomposition that affects system performance, scalability, and maintainability \citeP{nunes2019_transactional_contexts, faria2022_code_vectorization_microservices, rathod2023_industry4_refactoring}. Nunes \textit{et al.} \citeP{nunes2019_transactional_contexts} emphasize the need to focus on transactional contexts rather than structural dependencies when decomposing monolithic systems. Their approach aims to preserve business logic while enabling effective service extraction. Similarly, Rathod \textit{et al.} \citeP{rathod2023_industry4_refactoring} propose using Relational Topic Modeling (RTM) to combine structural dependencies with semantic analysis, ensuring that refactoring operations improve design quality.

\textbf{Potential Solution:} A hybrid approach combining ML techniques with domain-driven design and expert knowledge can refine results and ensure practical applicability. Increasing model explainability can also help software architects validate and refine service decomposition outcomes.

\subsection{Data Quality and Availability}
ML models rely on diverse datasets, including source code repositories, execution traces, logs, and performance metrics. However, data unavailability, inconsistency, incompleteness, and noise significantly impact the accuracy and reliability of ML-based approaches \citeP{zhang2022_deeptralog_combined_anomaly, daoud2020_microservice_identification, du2018_anomaly_container_microservices, kalia2020_mono2micro, nunes2019_transactional_contexts, saidi2022_structural_dependency_microservices, bajaj2024_gtmicro_nlp_microservices, sooksatra2022_vae_gnn_approach, liu2020_dual_clustering_microservices, liang2024_unsupervised_log_anomaly, shahini2024_autoencoder_anomaly_detection, zhang2022_putracead_trace_anomaly_detection, li2021_tracing_data_anomaly_detection, gan2019_seer_performance_debugging, mathai2021_hgnn_microservices_representation, song2024_autonomous_fault_classification, chen2022_tracegra_anomaly_detection, al_debagy2019_decomposition_method, chen2023_dynamic_static_features, xu2023_heterogeneous_failure_diagnosis, rathod2023_industry4_refactoring, yang2019_miras_resource_allocation_workflows, vera2021_microservices_backlog_genetic, trabelsi2024_micromatic_automation, shi2022_bsdg_anomaly_detection, qian2023_graph_clustering_extraction, wang2024_fault_detection_transformer, trabelsi2024_magnet_graph_nn, romani2022_data_centric_identification, trabelsi2023_type_based_microservices}. Mathai \textit{et al.} \citeP{mathai2021_hgnn_microservices_representation} highlight the dependency of decomposition techniques on external artifacts, such as runtime traces and commit histories, which are often incomplete or unavailable. Similarly, Bajaj \textit{et al.} \citeP{bajaj2024_gtmicro_nlp_microservices} stress that inconsistent SDLC artifacts, including use case models and functional requirements, pose challenges in greenfield developments. Beyond data quality issues, the lack of data also plays a critical role in shaping the type of machine learning techniques. The scarcity of large, labeled datasets likely limited the use of deep learning techniques, favoring classical ML and graph-based models instead.

\textbf{Potential Solution:} Robust data preprocessing pipelines, data augmentation techniques, and validation mechanisms can mitigate inconsistencies and improve reliability. Establishing standardized datasets for benchmarking ML models can also facilitate the development of more robust approaches.

\subsection{Scalability Concerns}
Scalability becomes a critical issue when ML models are applied to large and complex monolithic systems. These models often experience increased computational overhead, leading to performance bottlenecks and delays in the migration process \citeP{kamimura2018_microservice_candidates, zhang2022_deeptralog_combined_anomaly, chen2022_edge_attention_localization, sooksatra2022_vae_gnn_approach, al_debagy2021_dependencies_based_decomposition, liu2020_dual_clustering_microservices, sun2024_failure_localization_autoencoder, li2021_tracing_data_anomaly_detection, gan2019_seer_performance_debugging, cao2022_domain_oriented_decomposition, luan2024_resource_optimization, rathod2023_industry4_refactoring, song2023_chainsformer_latency_aware_provisioning, vera2021_microservices_backlog_genetic, lv2024_graph_rl_deployment, bajaj2020_partial_migration_cloud_native, santos2024_performance_forecast_microservices, shafi2024_cdascaler_autoscaling, daoud2021multi, chou2021_security_privacy_nn, ray2023_microservice_placement_edge, tong2023_gma_autoscaling_edge_cloud, li2022_score_resource_orchestration, zeng2023_topology_adaptive_provisioning, song2024_autonomous_fault_classification}. Tong \textit{et al.} \citeP{tong2023_gma_autoscaling_edge_cloud} discuss the inefficiencies of centralized ML strategies in edge-cloud environments, where distributed architectures complicate data collection and synchronization. These challenges highlight the need for efficient processing strategies to handle large-scale migration tasks.

\textbf{Potential Solution:} Scalable ML frameworks, distributed computing, and incremental learning approaches can enhance efficiency. Optimized data vectorization and clustering techniques can also reduce computational costs.

\subsection{Integration Complexities}
Integrating ML into migration workflows is complex due to the diversity of tools, platforms, and architectures involved \citeP{nunes2019_transactional_contexts, yang2019_miras_resource_allocation_workflows, lv2024_graph_rl_deployment, ray2023_microservice_placement_edge}. Lv \textit{et al.} \citeP{lv2024_graph_rl_deployment} introduce a Graph Convolutional Network (GCN) combined with deep reinforcement learning (DRL) to optimize microservice deployment. While this approach enhances decision-making, its integration into existing migration frameworks remains challenging. Ray \textit{et al.} \citeP{ray2023_microservice_placement_edge} explore ML-driven microservice placement strategies in edge computing, emphasizing the need for adaptive, automated decision-making processes.

\textbf{Potential Solution:} Developing modular ML components with standardized APIs can improve interoperability. Using containerization and microservices for ML components simplifies deployment and integration with existing architectures.

\subsection{Specialized Skills and Resource Requirements}
ML-based migration requires expertise in both machine learning and software engineering, which many development teams lack \citeP{cai2021_tracemodel_microservices, abdullah2019_autoscaling_policies, zhang2022_putracead_trace_anomaly_detection, li2021_tracing_data_anomaly_detection, gan2019_seer_performance_debugging, li2021_deepstitch_cross_layer, gan2022_practical_cloud_performance, khan2023_dynamic_resource_management, mathai2021_hgnn_microservices_representation, song2023_chainsformer_latency_aware_provisioning, yang2019_miras_resource_allocation_workflows, lv2024_graph_rl_deployment, dehghani2022_migration_model_driven_rl, trabelsi2024_micromatic_automation, santos2024_performance_forecast_microservices}. Santos \textit{et al.} \citeP{santos2024_performance_forecast_microservices} highlight the complexity of implementing ensemble learning models for performance forecasting, while Dehghani \textit{et al.} \citeP{dehghani2022_migration_model_driven_rl} emphasize the difficulty of applying AI-based migration techniques due to their reliance on reinforcement learning and neural networks.

\textbf{Potential Solution:} Combining specialized training for developers with thoughtfully designed black-box tools that abstract technical complexity while preserving essential controls, supported by cross-disciplinary collaboration between software engineers and ML practitioners. 

%% file: Sections/9-discussion.tex
\section{Discussion}
\label{Section: Discussion}

\begin{figure*}[ht]
\includegraphics[width=\linewidth]{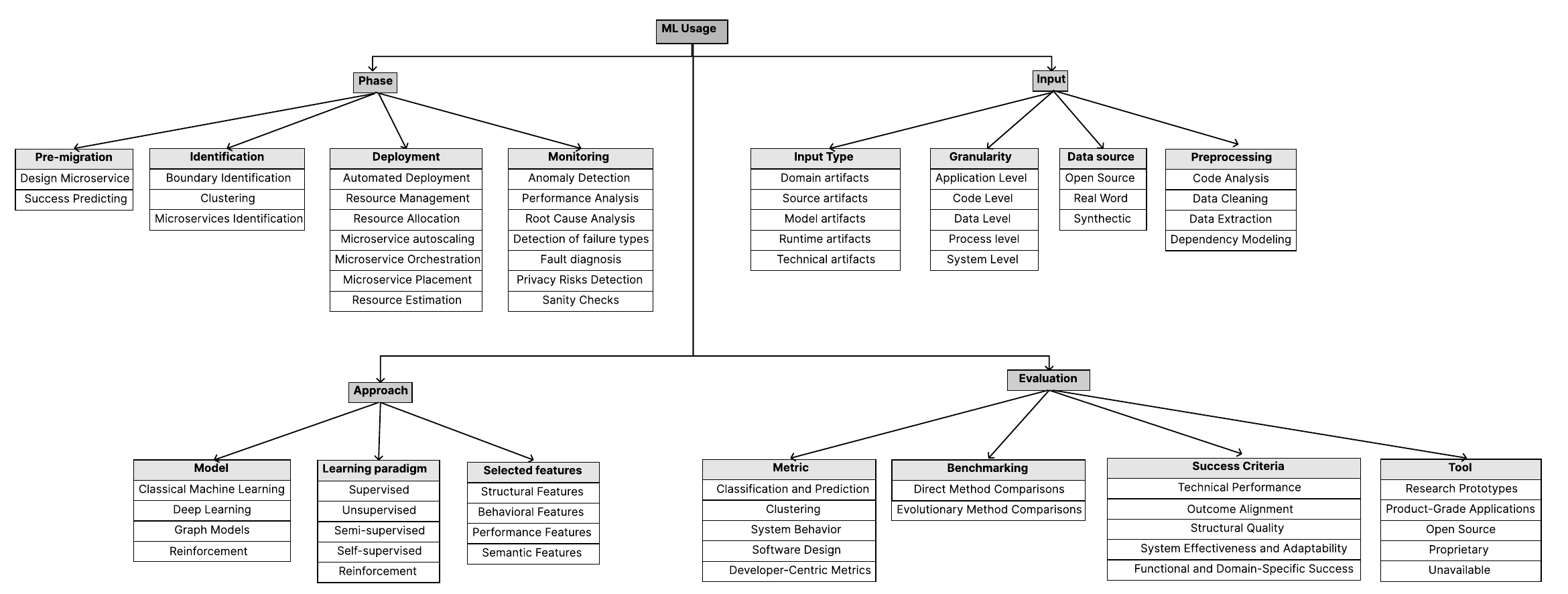}
\caption{\revised{ML Usage in Migration: A Classification Overview} }
\label{fig:taxOverview}
\end{figure*}

Figure~\ref{fig:taxOverview} presents a high-level classification derived from our answers to the research questions, summarising the key findings discussed in the previous sections. Building on this classification, we further analyse the four core dimensions addressed by our study, including Phases, Input Types, ML Approaches, and Evaluation Methods. In addition to examining each dimension independently, we conduct a cross-dimension analysis to reveal how these aspects co-occur across the selected primary studies. Finally, we synthesise these insights into concrete recommendations to support practitioners and researchers in effectively applying ML techniques during migration.

\subsection{Phases}

Our findings highlight that Identification is the phase in which ML approaches have been the most applied by researchers, with 39 PSs addressing automating this phase. It is followed by Monitoring (28 PSs) and Deployment (12 PSs), reflecting the research community's emphasis on defining microservices, monitoring their functionality, and ensuring their deployment into production environments. The pre-migration is perceived as straightforward or reliant on domain-specific knowledge, which limits its appeal to academic exploration. Also, the variability in business logic and technical landscapes across domains makes it challenging to generalise findings, further discouraging research. Additionally, we found no PSs proposing an ML-based approach to assist the Packaging phase. This phase remains largely overlooked, despite relying on extensive code generation and refactoring—both of which are still predominantly manual and time-consuming. Earlier machine learning approaches struggled to automate this phase effectively. However, the advent of Large Language Models (LLMs) presents new opportunities to address these challenges. LLMs have the potential to automate substantial aspects of the code generation process, such as API generation, offering solutions that are not only automated but also more consistent and accurate \cite{hou2023large}. \revised{This} shift highlights an opportunity for future research to explore the potential integration of LLMs into the packaging phase to assist automation efforts. While their actual effectiveness in this context remains to be empirically validated, recent successes of LLMs in tasks such as code generation, bug fixing, and software refactoring suggest that they may offer valuable capabilities for addressing migration-related challenges~\cite{Chen2023codellama, Ahmad2023context}

\subsection{Input}

The effectiveness of any machine learning-based approach, including those for microservices migration, depends on the quality, diversity, and preparation of its input data. The classification categorises inputs into five main types: domain artefacts (e.g., API documentation), Executable software model (e.g., source code), model artifacts (e.g., use cases), runtime artefacts (e.g., resource metrics, logs), and technical artifacts (e.g., Servers information). Among these, runtime artifacts are the most frequently utilised, appearing in 49 PSs, because they are crucial for capturing dynamic system behaviours, making them particularly valuable in monitoring and deployment phases. In contrast, domain and model artifacts are predominantly used during the identification and pre-migration phases.

Data sources vary, with most PSs relying on open-source data (50 PSs), due to their accessibility. Real-world data sources (28 PSs), and synthetic data sources (11 PSs) provide additional data, though their use is limited by accessibility and confidentiality challenges. The limited use of real-world and industrial systems justifies the gap in using academic findings on practical, large-scale applications. 

Preprocessing steps, including data extraction (39 PSs), data cleaning (24 PSs), code analysis (47 PSs), and dependency modelling (21 PSs), are used for preparing data for machine learning models. These steps ensure that raw data is transformed into a structured and meaningful format, enabling models to effectively analyse system components, dependencies, and relationships. While these preprocessing techniques have been explored in various studies, challenges such as handling incomplete or noisy data and automating complex code analysis remain critical areas for further research.

\subsection{Approaches}

\begin{figure}[ht]
\centering
\includegraphics[width=0.48\textwidth]{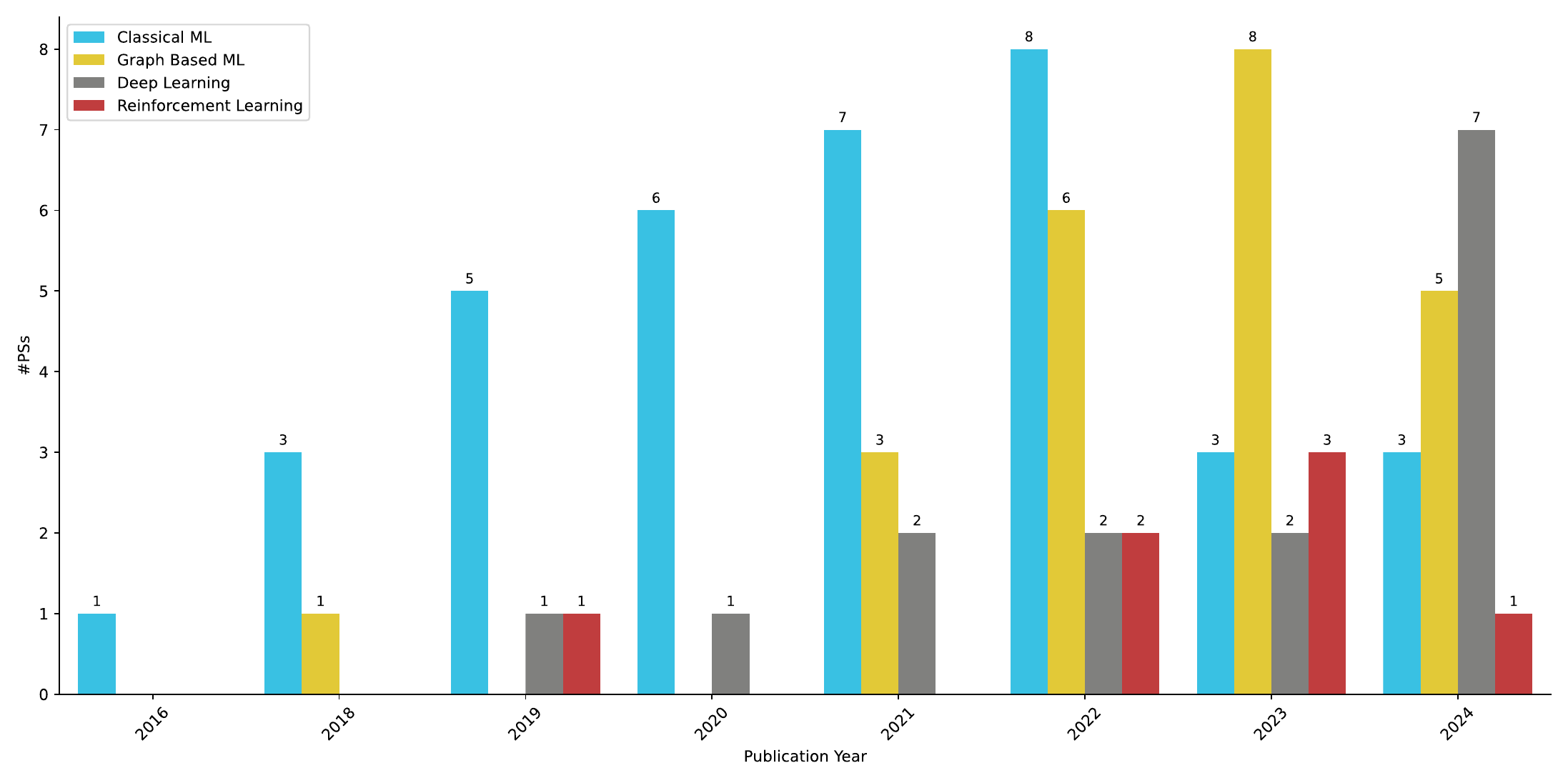}
\caption{Trends in ML Models usage across publication years}
\label{fig:MLperYear}
\end{figure}

The PSs employ a diverse range of ML models that we categorised into classical machine learning, deep learning, graph-based methods, and reinforcement learning. Among these, classical machine learning is the most prominent, appearing in 30 PSs, with clustering, classification, and regression techniques commonly used in the PSs. However, a cross-analysis of ML technique usage over time (see Figure~\ref{fig:MLperYear}) reveals a temporal trend: while classical ML dominated earlier works (before 022), recent PSs show increasing adoption of deep learning and graph-based approaches. Deep learning is used particularly in the monitoring \revised{phase} to model complex behaviors and predict root causes of failures, leveraging architectures such as autoencoders, transformers, and recurrent neural networks. Graph-based methods, used in 23 PSs, are particularly effective in capturing structural relationships between code components or log traces. Reinforcement learning appears in 19 PSs, mainly in deployment and monitoring phases, where it is applied for adaptive resource management and anomaly detection. Supervised learning is predominantly employed in monitoring, where labeled data is available. Unsupervised learning techniques are reported in 55 PSs, especially in the identification phase, where labeled datasets are typically unavailable. For instance, clustering algorithms like DBSCAN and K-Means are used in 6 PSs to identify microservices in monolithic architectures, with evaluation criteria focusing on cohesion and coupling.

ML models in these PSs leverage a variety of feature types to enhance different phases of the migration process. Performance features (24 PSs), such as CPU usage, memory consumption, and response time, are commonly used for performance prediction, anomaly detection, and deployment optimisation. Structural Features (51 PSs) enable ML models to learn structural and relational patterns, facilitating service decomposition, impact analysis, and dependency resolution. Semantic features (30 PSs), such as function names, API descriptions, and business logic embeddings, support domain-aware clustering and automated service identification by capturing functional similarities. Behavioural features (35 PSs), including invocation paths, transactional dependencies, and execution traces, provide insights into runtime behaviour, helping ML models understand component interactions and operational patterns. While these feature types enhance model effectiveness, challenges remain in automating feature extraction and handling noisy dependencies.

\subsection{Evaluation}

The evaluation of microservices migration approaches relies on a combination of metrics, benchmarking, and success criteria.
The analysis of evaluation metrics in the reviewed PSs highlights a strong emphasis on classification and prediction metrics (45 PSs), which are predominantly used to assess the accuracy and effectiveness of ML models in microservices identification and anomaly detection. Software design metrics (23 PSs) follow, focusing on cohesion, coupling, and modularity to ensure maintainability and architectural quality, exclusively used in the identification phase. System behaviour metrics (18 PSs), such as response time and resource \revised{utilisation}, are widely used to evaluate the performance and scalability of migrated microservices. Clustering metrics (12 PSs) provide insights into the quality of service decomposition, ensuring well-formed and meaningful microservices groupings. In contrast, developer-centric metrics (3 PSs) remain underexplored, despite their importance in validating the practicality and usability of automated migration approaches.

Benchmarking is a key validation method, with direct method comparisons being the most common (42 PSs), where approaches are evaluated against state-of-the-art techniques. Evolutionary method comparisons, which assess how approaches improve with new features, are significantly less frequent (6 PSs). Expanding evolutionary benchmarking could offer deeper insights into model robustness, scalability, and long-term adaptability. 

 Tool availability remains a major concern across both academic and industrial contexts. Among the 81 primary studies reviewed, only 21 provide open-source implementations, while 60 do not disclose any accessible tool, severely limiting reproducibility, evaluation, and practical adoption. To complement our literature analysis, we conducted a supplementary web search using the query "microservices migration" (tool OR platform OR product) (commercial OR enterprise OR SaaS). This search revealed that three commercial tools exist, notably \href{https://www.ibm.com/cloud/mono2micro}{IBM Mono2Micro}, \href{https://www.vfunction.com}{vFunction}, and \href{https://commercetools.com}{Commercetools Composable Commerce}, which support aspects of migration such as service boundary detection, observability-based decomposition, and incremental modernisation. However, these tools are commercially licensed, not open source, and do not explicitly leverage machine learning techniques. This reinforces our observation that ML-driven support for automated migration remains limited, non-reproducible, and poorly standardised. Future work should prioritise the development of publicly accessible tools and transparent ML-based solutions, along with standardised evaluation frameworks, to ensure both industrial relevance and scientific rigour.

\subsection{Cross-Dimensions Analysis}

\begin{figure*}[ht]
\center
 \includegraphics[width=1\linewidth]{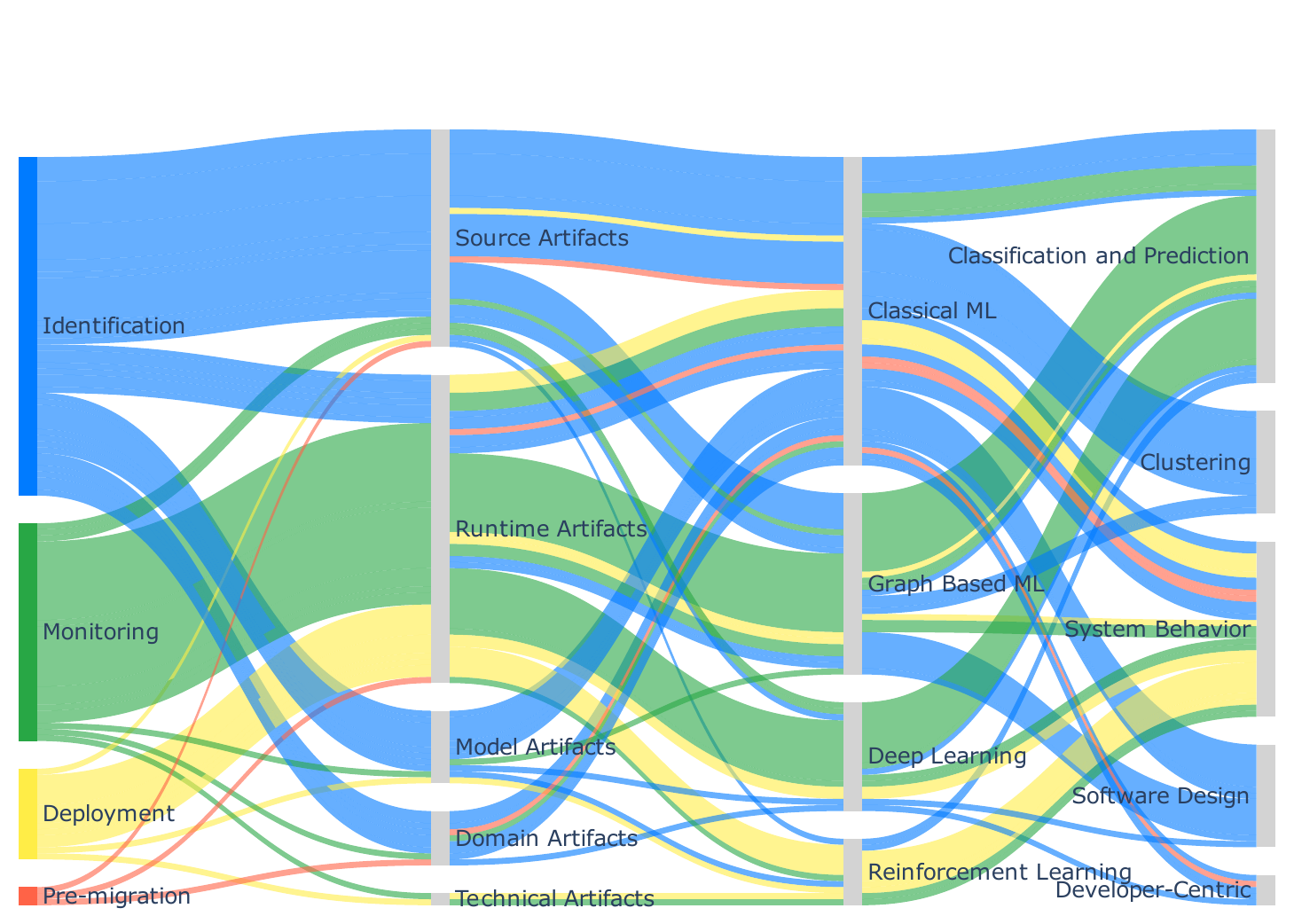}
\caption{Cross-dimensions analysis of ML-based migration approaches. Each stream represents a path linking Migration Phase, Input Type, ML Technique, and Evaluation Metric. }
 \label{fig:cross-rqs}
 \end{figure*}

While each research question focused on a specific dimension, we extended our analysis with a cross-dimensional examination to explore how migration phases, input types, ML models, and evaluation metrics co-occur across the selected primary studies. 
To support this analysis, we constructed a \textit{Sankey Diagram} (Figure~\ref{fig:cross-rqs}) that visualises complete migration paths derived from the selected primary studies. Each stream in the diagram corresponds to a unique combination of four key dimensions: Migration Phase $\rightarrow$ Input Type $\rightarrow$ ML Model $\rightarrow$ Evaluation Metric. The width of each stream is proportional to the number of PSs following that path, while the colour indicates the originating migration phase. An interactive version of this diagram is available online at \href{https://imen-trabelsi.github.io/SLR-MS-Migration/Cross-RQs-Interactive-Diagram.html}{this link}\footnotemark, where hover tooltips display the full path along with the corresponding number of supporting PSs.
\footnotetext{An interactive version of this figure is available at: \url{https://imen-trabelsi.github.io/SLR-MS-Migration/Cross-RQs-Interactive-Diagram.html}. Hovering over a stream reveals the complete path and the associated number of PSs.}

To structure our cross-dimensional analysis, we first identified the most frequent migration paths across all reviewed studies. These represent combinations of migration phase, input type, ML technique, and evaluation metric that occur most often. Table~\ref{tab:top-paths} presents the prominent paths reported in five or more primary studies.
 
\begin{table}[ht]
\centering
\caption{Most Frequent Migration Paths across Phase, Input, ML Technique, and Evaluation}
\label{tab:top-paths}
\resizebox{\linewidth}{!}{%
\begin{tabular}{|l|l|l|l|c|}
\hline
\textbf{Phase} & \textbf{Input Type} & \textbf{ML Technique} & \textbf{Evaluation Metric} & \textbf{\#PSs} \\
\hline
Monitoring & Runtime Artifacts & Graph Based ML & Classification and Prediction & 13 \\
Monitoring & Runtime Artifacts & Deep Learning & Classification and Prediction & 10 \\
Identification & Source Artifacts & Classical ML & Software Design & 7 \\
Identification & Source Artifacts & Classical ML & Clustering & 7 \\
Identification & Source Artifacts & Graph Based ML & Software Design & 6 \\
Deployment & Runtime Artifacts & Reinforcement Learning & System Behavior & 5 \\
\hline
\end{tabular}
}
\end{table}

This analysis highlights some dominant paths in how machine learning is applied across microservices migration phases. The most frequent migration paths emerge in the monitoring phase. These studies typically rely on runtime artifacts, such as logs, execution traces, and system-level metrics, and apply either graph-based ML or deep learning. Graph-based ML is used in 13 primary studies (\citeP{xu2023_heterogeneous_failure_diagnosis, chen2023_bert_htlg_detection, chen2022_edge_attention_localization, kong2024_fault_localization_span, shi2022_bsdg_anomaly_detection, gan2022_practical_cloud_performance, huang2023_twin_graph_detection, liang2024_unsupervised_log_anomaly, chen2022_tracegra_anomaly_detection, gan2021_sage_ml_performance_debugging, sun2024_failure_localization_autoencoder, zhang2022_putracead_trace_anomaly_detection, zhang2022_deeptralog_combined_anomaly}). Deep learning is used in 10 PSs (\citeP{ding2024_trace_anomaly_detection, luan2024_resource_optimization, gan2019_seer_performance_debugging, li2021_deepstitch_cross_layer, shahini2024_autoencoder_anomaly_detection, song2024_autonomous_fault_classification, wang2024_fault_detection_transformer, tan2024_maad_anomaly_detection, liu2020_anomaly_detection_microservices, cai2021_tracemodel_microservices}). Both configurations are primarily evaluated using classification and prediction metrics, such as precision, recall, F1-score, and AUC. These studies address anomaly detection, fault prediction, or behavioral monitoring in production environments.

In the identification phase, the dominant input remains source artifacts, including source code, call graphs, and code embeddings. Classical machine learning techniques such as K-Means, DBSCAN, and Agglomerative Clustering are applied in conjunction with either software design metrics 7 PSs (\citeP{trabelsi2023_type_based_microservices, nitin2022_cargo_dependency_analysis, al2021microservice, saidi2023_ddd_migration_microservices, sellami2022_hierarchical_dbscan_microservices, cao2022_domain_oriented_decomposition, faria2022_code_vectorization_microservices}) apply classification models to source code, or clustering to model-level inputs such as use-case diagrams and class diagrams.

In the deployment phase, most approaches rely on runtime artifacts and leverage reinforcement learning for dynamic optimisation, resource allocation, or autoscaling. These appear in 5 studies (\citeP{lv2024_graph_rl_deployment, yang2019_miras_resource_allocation_workflows, song2023_chainsformer_latency_aware_provisioning, ray2023_microservice_placement_edge, lv2022_deployment_edge_computing}), where the evaluation relies on system behavior metrics such as response time, memory consumption, or CPU utilisation. A few additional studies (e.g., \citeP{joseph2019_fuzzy_rl_allocation}) use classical ML or deep learning, but they are less common in this phase.

Several underexplored paths also emerged. These include configurations where domain-level artifacts (e.g., API documentation) are combined with classical ML and evaluated with system behavious metrics (e.g., \citeP{gysel2016_service_cutter_decomposition }) or cases where runtime input is introduced in the identification phase to guide service boundary prediction through behavioral insight (e.g., \citeP{morais2021_ontology_microservices }, \citeP{abdullah2019_unsupervised_web_decomposition }).

Beyond full migration paths, we examined pairwise relationships between individual dimensions to surface finer-grained patterns. For instance, input types vary significantly across phases. Source artifacts dominate the identification phase in 26 PSs, where they support architectural decomposition and service extraction (e.g.,  \citeP{trabelsi2023_type_based_microservices, desai2021_gnn_outlier_refactoring, zhong2023_spectral_clustering_industrial_legacy, nitin2022_cargo_dependency_analysis, rathod2023_industry4_refactoring}). Runtime artifacts are prevalent in both monitoring and deployment, used to inform operational decisions or detect abnormal behaviors \citeP{chou2021_security_privacy_nn, ding2024_trace_anomaly_detection , li2021_tracing_data_anomaly_detection}. Domain and model artifacts are primarily used in pre-migration or identification, reflecting their importance in early planning \citeP{alshammari2023_genetic_hpc_migration, daoud2020_microservice_identification_business }. Configuration artifacts appear almost exclusively in deployment studies that target environmental tuning or resource scheduling \citeP{ray2023_microservice_placement_edge, yang2019_miras_resource_allocation_workflows }.

Similarly, ML technique choices differ markedly across phases. Classical ML is most prominent in the identification phase, used for clustering and classification (e.g., \citeP{trabelsi2023_type_based_microservices, bajaj2020_partial_migration_cloud_native , morais2021_ontology_microservices, abdullah2019_unsupervised_web_decomposition }). Graph-based models also used for the identification phase, particularly when structural relations are central to the decomposition logic \citeP{trabelsi2024_magnet_graph_nn, nunes2019_transactional_contexts}. Approaches for monitoring, on the other hand, favor deep learning techniques that can model temporal or sequential patterns (e.g.,\citeP{chen2022_tracegra_anomaly_detection, 
gan2021_sage_ml_performance_debugging,
zhang2022_putracead_trace_anomaly_detection, 
tan2024_maad_anomaly_detection}). Reinforcement learning appears predominantly in the deployment phase, applied in contexts requiring adaptive policy learning \citeP{lv2024_graph_rl_deployment, song2023_chainsformer_latency_aware_provisioning,
lv2022_deployment_edge_computing,
yang2019_miras_resource_allocation_workflows,
ray2023_microservice_placement_edge}.

Evaluation practices reflect these distinctions. Studies in the identification phase rely mostly on software design metrics, such as modularity, cohesion, and coupling, which evaluate the structural quality of identified microservices candidates \citeP{qian2023_graph_clustering_extraction,
desai2021_gnn_outlier_refactoring,
zhong2023_spectral_clustering_industrial_legacy,
rathod2023_industry4_refactoring,
al2021microservice,
chen2023_dynamic_static_features,
liu2024_migration_graph_nn }. Most monitoring studies (17 out of 18 PSs) rely on classification and prediction metrics such as precision, recall, F1-score, and AUC.  However, \citeP{tong2023_gma_autoscaling_edge_cloud} validated their approach using only system behavior metrics, including Average Waiting Time (AWT), SLA violation rate, latency, and the standard deviation of autoscaling performance. Other studies \citeP{gan2022_practical_cloud_performance,
gan2021_sage_ml_performance_debugging,
luan2024_resource_optimization } combined classification metrics with system behavior metrics.
All deployment studies apply system behavior metrics such as latency or throughput to assess performance improvements and adaptability. Developer-centric or human-in-the-loop metrics are rare, particularly in pre-migration \citeP{alshammari2023_genetic_hpc_migration}, and during the identification phase \citeP{stojanovic2023application,
morais2021_ontology_microservices ,
gysel2016_service_cutter_decomposition }.

We also examined input–technique pairings. Source artifacts are mostly combined with classical ML and, to a lesser extent, graph-based ML \citeP{trabelsi2024_magnet_graph_nn,
eski2018_microservices_extraction,
kong2024_fault_localization_span}. Runtime artifacts drive the use of deep learning and reinforcement learning, given their rich behavioral data and need for temporal modeling \citeP{lv2024_graph_rl_deployment,
song2023_chainsformer_latency_aware_provisioning,
khan2023_dynamic_resource_management,
lv2022_deployment_edge_computing,
ding2024_trace_anomaly_detection,
chen2022_edge_attention_localization}. Domain artifacts are used with supervised techniques aimed at classification or prediction tasks during specially during the indentification phase \citeP{sun2022_expert_system_identification,
al_debagy2019_decomposition_method}.

\subsection{Recommendations for Researchers and Practitioners }

Based on the insights gathered, we provide recommendations to address the challenges of using machine learning during the migration from monolithic systems to microservices. 

\paragraph{\textbf{Recommendations for Researchers}} 

We present researchers with current challenges and future research directions to address key obstacles and enhance the effectiveness of ML-based migration approaches.

\textbf{Explore hybrid ML models} that offer a promising approach for improving microservice identification, deployment optimisation, and anomaly detection by leveraging the strengths of multiple learning paradigms.\\
\textbf{Rationale:} Findings from RQ4 emphasise the need for hybrid models to tackle the challenges of microservices migration. For example, graph-based ML models (e.g., GNNs) combined with clustering algorithms (e.g., DBSCAN, K-Means) can enhance microservice identification by capturing both structural dependencies and functional clustering. For deployment optimisation, integrating reinforcement learning (RL) with predictive models (e.g., LSTMs) can enable dynamic resource allocation and adaptive scaling based on workload patterns. Future research should focus on implementing hybrid ML techniques that enhance automation, adaptability, and interpretability in microservices migration, ensuring they effectively handle evolving architectures, dynamic workloads, and complex service interactions while maintaining high efficiency.

 \textbf{Address data accessibility issues} by collaborating with industry stakeholders to create anonymised datasets.\\
\textbf{Rationale:} Insights from RQ2 (Input) and RQ5 (Data Availability) underscore the importance of data accessibility and quality to ensure the effectiveness and reliability of ML approaches. Collaboration with industry stakeholders facilitates the creation of datasets that reflect real-world complexities while safeguarding privacy. Researchers should develop standardised anonymisation techniques, propose privacy-preserving data-sharing frameworks, and advocate for open-access benchmark datasets to improve reproducibility. Additionally, they should engage with industry partners to establish domain-specific data collection methodologies that ensure ML models are trained on representative and diverse datasets.

 \textbf{Standardise evaluation metrics} by developing universal  benchmarking standards and engaging with academic and industrial communities to define success criteria for migration.\\
\textbf{Rationale:} Standardised metrics ensure fairness, reproducibility, and comparability across research efforts, addressing inconsistencies in assessing ML-based migration approaches. Insights from RQ4 (Evaluation) and RQ5 (Evaluation Challenges) emphasize the necessity of well-defined benchmarks to advance the field effectively.

 \textbf{Explore the potential of ML in underexplored migration phases}, including both pre-migration and packaging. While our findings (RQ1) show that most ML-based approaches focus on identification, deployment, and monitoring, we observed significantly fewer efforts targeting the earlier phases, particularly business process analysis and modernisation, or the intermediate steps like code refactoring and service packaging. \\
\textbf{Rational:} Targeted research in these areas could provide tangible benefits, especially where current approaches rely heavily on manual expertise. For instance, the use of ML and AI techniques (e.g., NLP, process mining, or large language models) to support automated business process understanding and transformation could bridge the gap between legacy processes and microservice-oriented design. Similarly, leveraging ML to automate packaging and infrastructure preparation (e.g., configuration, API scaffolding) could reduce migration cost and risk. We encourage future work to investigate these directions where ML’s strengths (pattern discovery, learning from examples, code synthesis) align with clearly defined and repetitive tasks.

\paragraph{\textbf{Recommendations for Practitioners}} From the insights we collected, we provide practical guidance to help practitioners overcome key challenges and optimise the application of ML techniques during migration.

 \textbf{Adopt hybrid approaches} that combine machine-learning techniques with business elements to address critical challenges in microservice migration. \\
\textbf{Rationale:} Drawing from our findings in RQ4 and RQ5, hybrid approaches are essential for resolving challenges such as defining microservice boundaries, determining appropriate service sizes, and selecting optimal deployment resources. These approaches integrate the strengths of machine-learning techniques with business elements, enabling a balanced resolution of technical and business challenges. By aligning ML techniques usage with specific business objectives through iterative workflows and cross-disciplinary discussions, organisations can improve the accuracy and relevance of the proposed approach. Adaptive frameworks, incorporating user feedback and domain knowledge, further ensure deployment optimisation, cost-efficiency, and alignment with organisational goals.

 \textbf{Invest in data quality} by prioritising high-quality datasets and establishing workflows for preprocessing and validation.\\
\textbf{Rationale:} Findings from RQ2 and RQ5 underscore that high-quality data is critical for the reliability and effectiveness of ML models during migration phases. Practical workflows for validation and robust preprocessing in datasets. Utilising effective tools for cleaning and structuring data ensures its usability, while collaboration with domain experts aligns datasets with business objectives. This alignment improves the reliability of machine-learning techniques and ensures that the outcomes meet organisational needs. 

\textbf{Adapt scalable ML approaches} to address the complexities of microservices migration.\\
\textbf{Rationale:} Findings from RQ4 and RQ5 emphasise that scalable ML approaches are essential for managing the increasing complexities of microservices migration as software systems are getting larger. Distributed computing techniques enable efficient processing of large-scale systems, reducing processing time and improving the scalability of the proposed approach. Monitoring tools can track resource usage, predict workload trends, and dynamically optimise resource allocation, ensuring efficient resource management during deployment. By aligning these approaches with microservices architecture constraints, organisations can minimise downtime, reduce operational bottlenecks, and ensure scalability to more complex systems.

 \textbf{Upskill teams} through continuous learning initiatives tailored to both machine-learning and software engineers\\
\textbf{Rationale:} Findings from RQ5 highlight that successful microservices migration requires teams to possess a blend of specialised skills and collaborative expertise. Comprehensive learning initiatives, such as hands-on workshops, scenario-based case studies, and real-world projects, bridge the gap between theoretical knowledge and practical application. Fostering cross-functional collaboration among domain experts, ML engineers, and software architects ensures seamless teamwork and alignment. By keeping teams updated with the latest technological advancements, organisations can empower them to tackle both technical and collaborative challenges effectively, driving successful migration outcomes.


%% file: Sections/9.1-Threats2Validity.tex
\section{Threats to Validity}
\label{sec:Threats2Validity}
In this section, we identify potential threats associated with construct validity, internal validity, external validity, and conclusion validity.
\paragraph{Construct Validity} This aspect focuses on the sources used and the approach adopted for data collection. It includes the selection process of primary studies and the methodology employed to extract data in relation to the research questions.
\textbf{\emph{Exclusion of relevant studies}:} The possibility of overlooking relevant studies poses a threat to the study. Since our search was based on titles and keywords, there is a chance that some relevant studies were unintentionally excluded. The effectiveness of our search depends largely on how well digital libraries index research papers. To address this issue, we used seven widely recognized and comprehensive databases for literature reviews \cite{dyba2007applying}. Additionally, we conducted four rounds of snowballing to identify further potentially relevant studies.\\
\textbf{\emph{Selection bias in PSs}:} The process of selecting PSs may have led to the exclusion of relevant studies. As the selection was conducted manually, subjective judgment could have influenced the final set of chosen studies. To mitigate this risk, we clearly defined the study’s objectives and research questions beforehand, adhering to the PRISMA guidelines. We also established well-defined inclusion and exclusion criteria.\\
\textbf{\emph{Bias in data extraction}:} Since data extraction was performed manually, there is a risk of personal bias affecting the results. To reduce this risk, we designed a structured data collection form. Two researchers independently carried out the extraction process following the predefined form. We ensured consistency through interrater agreement and held multiple discussions involving all authors to reach a consensus. Furthermore, inconsistencies in terminology across PSs posed another challenge. We addressed this by systematically discussing and agreeing upon a standardized vocabulary.\\
\textbf{\emph{Scope Boundary Definitions}:} Our study exclusively examined ML-based approaches. Consequently, we excluded approaches that do not explicitly involve ML techniques—such as those based solely on process mining or architectural heuristics. This restricted scope enabled a focused analysis of the specific characteristics and limitations of ML techniques. However, we acknowledge that non-ML approaches, including process mining methods (e.g., \cite{baresi2017microservices}), can play a complementary role. In particular, process mining can generate valuable artifacts such as event logs or execution traces, which may serve as input to ML models. Future research could beneficially explore such hybrid combinations, capitalizing on the complementary strengths of process-oriented and learning-based techniques.

\paragraph{Internal Validity} Internal validity pertains to the methods employed in this study and the accuracy of the conclusions derived from them.\\
\textbf{ \emph{Review completeness}: }This study primarily investigates the use of ML techniques in the migration process: phases, input, approach, evaluation, and challenges. Some PSs may not provide exhaustive details on all these aspects. To minimize this limitation, we included only those studies that can answer at least three RQs.\\
\textbf{\emph{Research methodology}:} The methodology used in this study may introduce certain biases, potentially impacting the reliability and validity of the findings. To mitigate this risk, we followed the updated PRISMA guidelines and carefully curated our study selection process to ensure relevance. The inclusion and exclusion criteria were rigorously defined and reviewed by all authors to enhance objectivity.

\paragraph{External Validity} This concerns the extent to which our findings can be generalized across all migration approaches that use ML. Our review focuses exclusively on academic literature, meaning that industry practices may not be fully represented if they were not documented in research papers. Additionally, we only considered studies published within a specific timeframe, which could limit the generalizability of our conclusions. Nonetheless, we plan to conduct an industry-focused study to supplement these findings.

\paragraph{Conclusion Validity} This aspect pertains to the soundness of the conclusions drawn from the extracted data. To ensure validity, we conducted a thorough analysis of the extracted data and engaged in multiple discussion sessions to cross-verify our conclusions. We strictly based our findings on the data derived from the selected PSs, ensuring that our conclusions remain well-supported and justified.

%% file: Sections/10-conclusion.tex
\section{Conclusion}
\label{Section: Conclusion}
This systematic literature review examines the role of machine learning in automating the migration from monolithic systems to microservices. Using a PRISMA-based methodology, we analyzed 81 primary studies to understand the automated phases, the types of inputs used, the ML techniques applied, the evaluation methods used, and the challenges encountered.
Our findings indicate that ML-driven migration studies have primarily focused on service identification, monitoring, and deployment, while pre-migration analysis and microservice packaging remain largely unexplored. In particular, automating code generation and packaging tasks has received limited attention, despite the potential of emerging technologies such as Large Language Models. Additionally, runtime artifacts serve as the dominant data source, yet the scarcity of real-world datasets raises concerns about the practical applicability of ML techniques. Unsupervised learning remains the most common approach, particularly for service identification and anomaly detection, but evaluation practices vary widely, with system adaptability and real-world validation often overlooked. Despite ML’s potential to automate migration tasks, key challenges persist, including data availability, scalability, tool support, and the absence of standardized benchmarking. Addressing these challenges requires further research on underexplored migration phases, particularly pre-migration planning and packaging. Enhancing data accessibility through industry collaboration, privacy-preserving data-sharing frameworks, and the development of benchmark datasets will be crucial. Furthermore, exploring hybrid ML approaches that integrate multiple learning paradigms could improve accuracy and adaptability in migration processes. By addressing these gaps, future research can contribute to more effective and scalable ML-driven migration solutions.

%% file: Sections/appendices.tex
\appendix
\section*{Methodology Details} \label{app:methodology}
\subsection*{Inclusion Criteria and Exclusion Criteria}
\begin{table*}[h!]
\centering
\caption{ Examples of Excluded Studies Based on Specific Criteria}
\label{tab:exclusion-examples}
\begin{tabular}{|p{4cm}|p{2cm}|p{7.5cm}|}
\hline
\textbf{Criterion} & \textbf{Excluded Study} & \textbf{Justification} \\
\hline
IC7: The study provides enough information to answer at least 3 RQs & \cite{djogic2018monolithic} & The study provided only a high-level summary of its approach, without sufficient technical or evaluation detail. Only two research (RQ1 and RQ2) questions could be addressed based on the available content. \\
\hline
IC8: The study has its full text available online & \cite{Idris2022nlpBased}& The publication metadata was available, but access to the full article text was missing, preventing proper assessment. \\
\hline
IC9: The study provides sufficient migration details & \cite{fan2017migrating}& The paper outlines decomposition patterns and incremental migration steps but omits how evaluated and ML techniques used \\
\hline
IC10: The study uses an automated or semi-automated migration approach & \cite{faustino2024stepwise} & The study proposed architectural guidelines but did not implement or evaluate any tool or algorithm to automate the migration process. \\
\hline
EC5: The study does not provide enough details & \cite{fan2017migrating} & The paper briefly mentioned the use of ML for migration but did not describe the technique, inputs, or outputs, making it unsuitable for analysis. \\
\hline
EC6: The study does not provide an automated or semi-automated migration approach &\cite{faustino2024stepwise}  & The migration approach was based entirely on manual analysis and refactoring decisions by software architects, with no automation. \\
\hline
\end{tabular}
\end{table*}

\subsection*{Primary Studies (PSs) list}
Table \ref{tab:overview_selected_literature} show the list of the selected primary studies.
\begin{table*}[!htbp]
\caption{List of Selected PSs}
\label{tab:overview_selected_literature}
\centering
\footnotesize
\begin{threeparttable}
\scalebox{0.9}{ 
\begin{tabularx}{\linewidth}[t]{
    >{\raggedright\arraybackslash\hsize=2.7\hsize}X
    >{\centering\arraybackslash\hsize=0.1\hsize}X 
    >{\centering\arraybackslash\hsize=0.2\hsize}X
  }
\toprule
\textbf{Primary Study (PS)} & \textbf{Phase} & \textbf{Ref.} \\ \midrule
A DDD Approach Towards Automatic Migration to Microservices &I &
  \citeP{saidi2023_ddd_migration_microservices} \\
A Hierarchical DBSCAN Method for Extracting Microservices from Monolithic Applications &I&
  \citeP{sellami2022_hierarchical_dbscan_microservices} \\
A microservice decomposition method using a distributed representation of source code &I&
  \citeP{al2021microservice} \\
A Microservices Identification Method Based on Spectral Clustering for Industrial Legacy Systems &I&
  \citeP{zhong2023_spectral_clustering_industrial_legacy} \\
A multi-model based microservices identification approach &I&
  \citeP{daoud2021multi} \\
A New Decomposition Method for Designing Microservices &I&
  \citeP{al_debagy2019_decomposition_method} \\
An Automatic Extraction Approach: Transition to Microservices Architecture from Monolithic Application &I&
  \citeP{eski2018_microservices_extraction} \\
Anomaly Detection and Diagnosis for Container-Based Microservices with Performance Monitoring &M&
  \citeP{du2018_anomaly_container_microservices} \\
Autoencoder-Based Anomaly Detection in Microservices Using Distributed Tracing &M&
  \citeP{shahini2024_autoencoder_anomaly_detection} \\
Automatic Microservices Identification Across Structural Dependency &I&
  \citeP{saidi2022_structural_dependency_microservices} \\
Automatic Microservices Identification from a Set of Business Processes &I&
  \citeP{daoud2020_microservice_identification} \\
Automatic Migration-Enabled Dynamic Resource Management for Containerized Workload &D&
  \citeP{khan2023_dynamic_resource_management} \\
Autonomous Selection of the Fault Classification Models for Diagnosing Microservice Applications &M&
  \citeP{song2024_autonomous_fault_classification} \\
BertHTLG: Graph-Based Microservice Anomaly Detection Through Sentence-BERT Enhancement &M&
  \citeP{chen2023_bert_htlg_detection} \\
BSDG: Anomaly Detection of Microservice Trace Based on Dual Graph Convolutional Neural Network &M&
  \citeP{shi2022_bsdg_anomaly_detection} \\
CARGO: AI-Guided Dependency Analysis for Migrating Monolithic Applications to Microservices Architecture &I&
  \citeP{nitin2022_cargo_dependency_analysis} \\
Cdascaler: A Cost-Effective Dynamic Autoscaling Approach for Containerized Microservices &D&
  \citeP{shafi2024_cdascaler_autoscaling} \\
ChainsFormer: A Chain Latency-Aware Resource Provisioning Approach for Microservices Cluster &D&
  \citeP{song2023_chainsformer_latency_aware_provisioning} \\
Code Vectorization and Sequence of Accesses Strategies for Monolith Microservices Identification &I&
  \citeP{faria2022_code_vectorization_microservices} \\
DeepRest: Deep Resource Estimation for Interactive Microservices &D&
  \citeP{chow2022_deeprest_resource_estimation} \\
Deepstitch: Deep Learning for Cross-Layer Stitching in Microservices &M&
  \citeP{li2021_deepstitch_cross_layer} \\
DeepTraLog: Trace-Log Combined Microservice Anomaly Detection through Graph-based Deep Learning &M&
  \citeP{zhang2022_deeptralog_combined_anomaly} \\
Dependencies-Based Microservices Decomposition Method &I&
  \citeP{al_debagy2021_dependencies_based_decomposition} \\
Detecting Security and Privacy Risks in Microservices End-to-End Communication Using Neural Networks &M&
  \citeP{chou2021_security_privacy_nn} \\
Dynamic and Static Feature-Aware Microservices Decomposition via Graph Neural Networks &I&
  \citeP{chen2023_dynamic_static_features} \\
Enabling Practical Cloud Performance Debugging with Unsupervised Learning &M&
  \citeP{gan2022_practical_cloud_performance} \\
Enhancing Fault Localization in Microservices Systems Through Span-Level Using Graph Convolutional Networks &M&
  \citeP{kong2024_fault_localization_span} \\
Expert System for Automatic Microservices Identification Using API Similarity Graph &I&
  \citeP{sun2022_expert_system_identification} \\
Extracting Candidates of Microservices from Monolithic Application Code &I&
  \citeP{kamimura2018_microservice_candidates} \\
Facilitating the Migration to the Microservice Architecture via Model-Driven Reverse Engineering and RL &I&
  \citeP{dehghani2022_migration_model_driven_rl} \\
From a Monolith to a Microservices Architecture: An Approach Based on Transactional Contexts &I&
  \citeP{nunes2019_transactional_contexts} \\
From Legacy2Microservices: A Type-Based Approach for Microservices Identification Using ML and Semantic Analysis &I&
  \citeP{trabelsi2023_type_based_microservices} \\
From Monolithic Architecture Style to Microservice One Based on a Semi-Automatic Approach &I&
  \citeP{selmadji2020_transition_microservices} \\
Fuzzy Reinforcement Learning Based Microservice Allocation in Cloud Computing Environments &D&
  \citeP{joseph2019_fuzzy_rl_allocation} \\
GMA: Graph Multi-Agent Microservice Autoscaling Algorithm in Edge-Cloud Environment &M&
  \citeP{tong2023_gma_autoscaling_edge_cloud} \\
Graph Neural Network to Dilute Outliers for Refactoring Monolith Application &I&
  \citeP{desai2021_gnn_outlier_refactoring} \\
Graph-Reinforcement-Learning-Based Dependency-Aware Microservice Deployment in Edge Computing &D&
  \citeP{lv2024_graph_rl_deployment} \\
GTMicro—Microservice Identification Approach Based on Deep NLP Transformer Model for Greenfield Developments &I&
  \citeP{bajaj2024_gtmicro_nlp_microservices} \\
Heterogeneous Data-Driven Failure Diagnosis for Microservice Industrial Clouds Towards Consumer Digital Ecosystems &M&
  \citeP{xu2023_heterogeneous_failure_diagnosis} \\
High-Performance Computing-Enabled Probabilistic Framework for Migration from Monolith2Microservices Using GAs&P&
  \citeP{alshammari2023_genetic_hpc_migration} \\
Implementation of Domain-Oriented Microservices Decomposition Based on Node-Attributed Network &I&
  \citeP{cao2022_domain_oriented_decomposition} \\
Improving Industry 4.0 Readiness: Monolith Application Refactoring Using Graph Attention Networks &I&
  \citeP{rathod2023_industry4_refactoring} \\
Interpretable Failure Localization for Microservice Systems Based on Graph Autoencoder &M&
  \citeP{sun2024_failure_localization_autoencoder} \\
Learning Predictive Autoscaling Policies for Cloud-Hosted Microservices Using Trace-Driven Modeling &D&
  \citeP{abdullah2019_autoscaling_policies} \\
Learning-Based Microservice Placement and Migration for Multi-Access Edge Computing &D&
  \citeP{ray2023_microservice_placement_edge} \\
MAAD: A Distributed Anomaly Detection Architecture for Microservices Systems &M&
  \citeP{tan2024_maad_anomaly_detection} \\
Magnet: Method-Based Approach Using Graph Neural Network for Microservices Identification &I&
  \citeP{trabelsi2024_magnet_graph_nn} \\
Method of Microservices Division for Complex Business Management System Based on Dual Clustering &I&
  \citeP{liu2020_dual_clustering_microservices} \\
Microegrcl: An Edge-Attention-Based Graph Neural Network Approach for Root Cause Localization in Microservice Systems &M&
  \citeP{chen2022_edge_attention_localization} \\
MicroMatic: Fully Automated Microservices Identification Approach From Monolithic Systems &I&
  \citeP{trabelsi2024_micromatic_automation} \\
Microservice Anomaly Detection Based on Tracing Data Using Semi-Supervised Learning &M&
  \citeP{li2021_tracing_data_anomaly_detection} \\
Microservice Deployment in Edge Computing Based on Deep Q Learning &D&
  \citeP{lv2022_deployment_edge_computing} \\
Microservice Extraction Using Graph Deep Clustering Based on Dual View Fusion &I&
  \citeP{qian2023_graph_clustering_extraction} \\
Microservices Backlog: Genetic Programming Technique for Identification and Evaluation of Microservices From User Stories&I&
  \citeP{vera2021_microservices_backlog_genetic} \\
Microservices Performance Forecast Using Dynamic Multiple Predictor Systems &M&
  \citeP{santos2024_performance_forecast_microservices} \\
Migrating Monolith System to Microservices with Directed Graph Attention Neural Network &I&
  \citeP{liu2024_migration_graph_nn} \\
Minimize Cost for Containerized Microservices Under SLO via ML-Enhanced Layered Queuing Network Optimization &M&
  \citeP{luan2024_resource_optimization} \\
MIRAS: Model-Based Reinforcement Learning for Microservice Resource Allocation Over Scientific Workflows &D&
  \citeP{yang2019_miras_resource_allocation_workflows} \\
Mono2Micro: An AI-Based Toolchain for Evolving Monolithic Enterprise Applications to a Microservice Architecture &I&
  \citeP{kalia2020_mono2micro} \\
Monolith to Microservices: Representing Application Software Through Heterogeneous Graph Neural Network &I&
  \citeP{mathai2021_hgnn_microservices_representation} \\
Monolith to Microservices: VAE-Based GNN Approach with Duplication Consideration &I&
  \citeP{sooksatra2022_vae_gnn_approach} \\
Multilayered Fault Detection and Localization With Transformer for Microservice Systems &M&
  \citeP{wang2024_fault_detection_transformer} \\
Partial Migration for Re-Architecting a Cloud Native Monolithic Application into Microservices and FaaS &I&
  \citeP{bajaj2020_partial_migration_cloud_native} \\
PUTraceAD: Trace Anomaly Detection with Partial Labels Based on GNN and PU Learning &M&
  \citeP{zhang2022_putracead_trace_anomaly_detection} \\
Sage: Practical and Scalable ML-Driven Performance Debugging in Microservices &M&
  \citeP{gan2021_sage_ml_performance_debugging} \\
SCORE: A Resource-Efficient Microservice Orchestration Model Based on Spectral Clustering in Edge Computing &D&
  \citeP{li2022_score_resource_orchestration} \\
Seer: Leveraging Big Data to Navigate the Complexity of Performance Debugging in Cloud Microservices &M&
  \citeP{gan2019_seer_performance_debugging} \\
Service Cutter: A Systematic Approach to Service Decomposition &I&
  \citeP{gysel2016_service_cutter_decomposition} \\
The Application of  ChatGPT for Identification of Microservices &I&
  \citeP{stojanovic2023application} \\
Topology-Aware Self-Adaptive Resource Provisioning for Microservices &D&
  \citeP{zeng2023_topology_adaptive_provisioning} \\
Towards an Automatic Identification of Microservices from Business Processes &I&
  \citeP{daoud2020_microservice_identification_business} \\
Towards an Ontology-Driven Approach to Model and Analyze Microservices Architectures &I&
  \citeP{morais2021_ontology_microservices} \\
Towards Migrating Legacy2Microservice-Based Architectures: A Data-Centric Process for Microservice Identification &I&
  \citeP{romani2022_data_centric_identification} \\
Trace Anomaly Detection for Microservice Systems via Graph-Based Semi-Supervised Learning &M&
  \citeP{ding2024_trace_anomaly_detection} \\
Tracegra: A Trace-Based Anomaly Detection for Microservice Using Graph Deep Learning &M&
  \citeP{chen2022_tracegra_anomaly_detection} \\
TraceModel: An Automatic Anomaly Detection and Root Cause Localization Framework for Microservice Systems &M&
  \citeP{cai2021_tracemodel_microservices} \\
Twin Graph-Based Anomaly Detection via Attentive Multi-Modal Learning for Microservice Systems &M&
  \citeP{huang2023_twin_graph_detection} \\
Unsupervised Detection of Microservice Trace Anomalies through Service-Level Deep Bayesian Networks &M&
  \citeP{liu2020_anomaly_detection_microservices} \\
Unsupervised Learning Approach for Web Application Auto-Decomposition into Microservices &I&
  \citeP{abdullah2019_unsupervised_web_decomposition} \\
Unsupervised Microservice Log Anomaly Detection Method Based on Graph Neural Network &M&
  \citeP{liang2024_unsupervised_log_anomaly} \\
Visualization Tool for Designing Microservices with the Monolith-First Approach &P&
  \citeP{nakazawa2018_visualization_tool} \\

\bottomrule
\end{tabularx}
} 
\begin{tablenotes}
\scriptsize \centering
    \item[\ding{90}] \textbf{P}: Pre-migration;  \textbf{I}: Identification;  \textbf{D}: Deployment; \textbf{M}: Monitoring.
\end{tablenotes}
\end{threeparttable}
\end{table*}

\subsection*{Databases for Selected PSs}
\begin{figure}[ht]
\centering
 \includegraphics[width=0.7\linewidth]{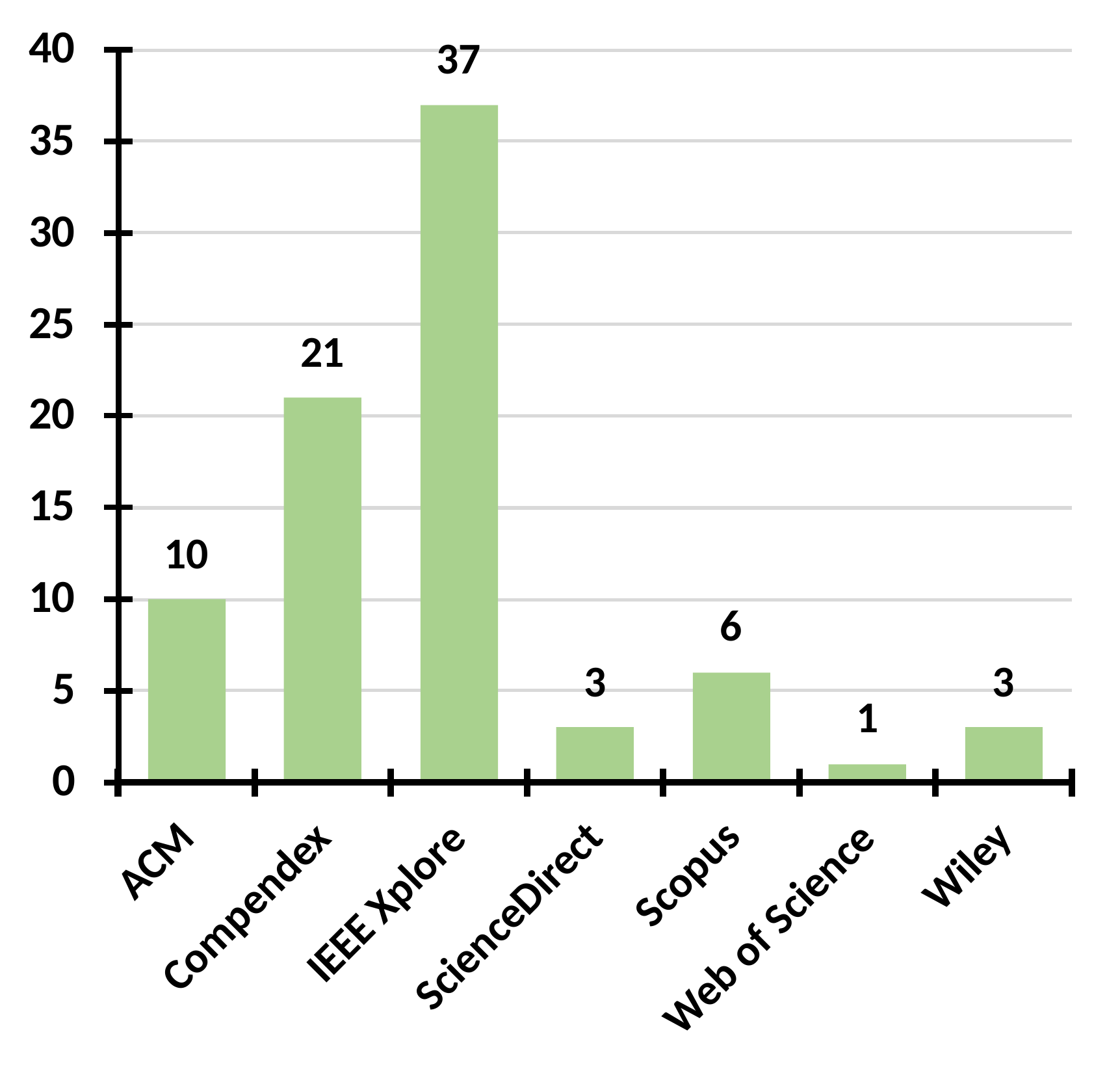}
 \caption{Databases for Selected PSs} 
 \label{fig:database_venues_for_PSs}
 \end{figure}

  Figure \ref{fig:database_venues_for_PSs} shows the databases for selected PSs. 

\section*{Definitions of Key Terms for RQ1} 
\label{app:definitions_phases}

\subsection*{Migration Phases}
\begin{itemize}[noitemsep,topsep=0pt,leftmargin=*]
    \item \textbf{Pre-migration}: Focuses on evaluating the current approaches, methods, and tools used in the legacy system. This phase also involves planning the migration by identifying suitable strategies, defining objectives, and outlining the steps needed to ensure a smooth transition.
    \item \textbf{Identification}: Defines the boundaries of prospective microservices. This involves detecting functional modules, mapping dependencies, and clustering components with business requirements.
    \item \textbf{Packaging}: Encapsulates the identified components into functional microservices. This phase includes defining service interfaces, managing dependencies, and generating missing components.
    \item \textbf{Deployment}: Implements and configures the necessary runtime infrastructure to host microservices, including containerization, service orchestration, network configuration, and integration with legacy or third-party systems.
    \item \textbf{Monitoring}: Ensures the reliability and efficiency of the microservices-based system post-migration. This phase involves continuous performance tracking, anomaly detection, and dynamic resource management to meet changing workload demands.
\end{itemize}

\subsection*{Automated Tasks}
\begin{itemize}[noitemsep,topsep=0pt,leftmargin=*]
    \item \textbf{Pre-migration Tasks}:
    \begin{itemize}[noitemsep,topsep=0pt,leftmargin=1.5em]
        \item Designing microservices: Assists in planning microservice architecture by analyzing legacy systems.
        \item Predicting success rates of migration: Uses predictive models to evaluate potential migration outcomes.
    \end{itemize}
    
    \item \textbf{Identification Tasks}:
    \begin{itemize}[noitemsep,topsep=0pt,leftmargin=1.5em]
        \item Boundary identification: Detects functional boundaries within monolithic systems.
        \item Clustering: Groups related components or services based on structural, semantic, or performance similarity using unsupervised learning.
        \item Microservices identification: Automates the process of determining candidate microservices.
    \end{itemize}

    \item \textbf{Deployment Tasks}:
    \begin{itemize}[noitemsep,topsep=0pt,leftmargin=1.5em]
        \item Automated deployment: Automates the deployment of microservices into production environments.
        \item Resource management: Allocates and monitors computing resources such as CPU, memory, and I/O bandwidth across microservices during runtime to optimize cost-efficiency and performance.
        \item Resource allocation: Dynamically allocates resources to microservices based on workload demands.
        \item Microservice autoscaling: Dynamically adjusts the number of microservice instances or resource allocations based on performance indicators such as CPU load, request latency, or throughput, in order to maintain service-level objectives and system responsiveness.
        \item Microservice orchestration: Manages dependencies and interactions between microservices.
        \item Microservice placement: Decides optimal placements for microservices within the infrastructure.
        \item Resource estimation: Estimates the resource needs of microservices to optimize deployment strategies.
    \end{itemize}

    \item \textbf{Monitoring Tasks}:
    \begin{itemize}[noitemsep,topsep=0pt,leftmargin=1.5em]
        \item Anomaly detection: Identifies abnormal behaviors or performance issues in microservices.
        \item Performance analysis: Evaluates the performance metrics of deployed microservices.
        \item Detection of failure types: Classifies failure types to facilitate troubleshooting.
        \item Fault diagnosis: Diagnoses root causes of failures to enable faster resolution.
        \item Root cause analysis: Pinpoints the underlying issues causing anomalies or failures.
        \item Privacy risks detection: Identifies potential privacy concerns in microservice interactions.
        \item Sanity check: Automated validations to confirm system states and outputs remain within expected bounds.
    \end{itemize}
\end{itemize}

\section*{Definitions of Key Terms for RQ2}  
\label{app:definitions_rq2}

\subsection*{Input Types}
\begin{itemize}[noitemsep,topsep=0pt,leftmargin=*]
    \item \textbf{Domain Artifacts}: High-level inputs that capture business and functional requirements of the system. These include:
    \begin{itemize}[noitemsep,topsep=0pt,leftmargin=1.5em]
        \item \textbf{API Documentation}: Details about available APIs and their interactions, often used to modularize microservices.
        \item \textbf{QoS Constraints}: Quality of Service parameters (e.g., response time, availability) that must be maintained during migration.
        \item \textbf{Architecture Recommendations}: Expert suggestions for migration strategies and component separation.
        \item \textbf{Functional Description}: High-level descriptions of system functionalities, often captured in ontologies.
    \end{itemize}

    \item \textbf{Runtime Artifacts}: Inputs derived from the system's runtime behavior, including:
    \begin{itemize}[noitemsep,topsep=0pt,leftmargin=1.5em]
        \item \textbf{Resource Metrics}: Data on resource consumption (e.g., CPU, memory usage).
        \item \textbf{Monitoring Metrics}: Data collected from system monitoring tools.
        \item \textbf{Performance Metrics}: Measurements of system performance (e.g., response time, latency).
        \item \textbf{Trace Logs}: Detailed records of system execution flow.
        \item \textbf{Workloads}: Descriptions of system usage scenarios and patterns.
    \end{itemize}

    \item \textbf{Model Artifacts}: Abstract representations of the system, including:
    \begin{itemize}[noitemsep,topsep=0pt,leftmargin=1.5em]
        \item \textbf{Business Processes}: Workflows and processes supported by the system.
        \item \textbf{UML Diagrams}: Visual representations of system design.
        \item \textbf{User Stories}: Descriptions of functionalities from the user's perspective.
    \end{itemize}

    \item \textbf{Source Artifacts}: Inputs related to the actual software and its configuration, including:
    \begin{itemize}[noitemsep,topsep=0pt,leftmargin=1.5em]
        \item \textbf{Source Code}: Implementation of the system, including modules and components.
        \item \textbf{Configuration Files}: Settings required for system operation.
        \item \textbf{Data Files}: Files used or generated by the system for operations.
    \end{itemize}

    \item \textbf{Technical Artifacts}: Supporting information for system operation and migration planning, including:
    \begin{itemize}[noitemsep,topsep=0pt,leftmargin=1.5em]
        \item \textbf{Server Information}: Details about the infrastructure supporting the system.
    \end{itemize}
\end{itemize}

\subsection*{Input Granularity}
\begin{itemize}[noitemsep,topsep=0pt,leftmargin=*]
 \item \textbf{Application Level}: Examines inputs at a higher level of abstraction, focusing on files, APIs, and use cases.
    \begin{itemize}[noitemsep,topsep=0pt,leftmargin=1.5em]
        \item \textbf{Use Case}: Scenarios describing user interactions with the application.
        \item \textbf{Files}: Application files (e.g., configuration or resource files).
        \item \textbf{URI}: Resources or endpoints within the application.
        \item \textbf{API}: Application Programming Interfaces for interacting with functionalities.
    \end{itemize}

    \item \textbf{System Level}: Investigates components and interactions within the larger system architecture.
    \begin{itemize}[noitemsep,topsep=0pt,leftmargin=1.5em]
        \item \textbf{Services}: Functional units delivering specific capabilities.
        \item \textbf{Microservices}: Independent units designed for specific tasks.
        \item \textbf{Node}: Individual units in distributed systems (e.g., servers, containers).
        \item \textbf{Component}: A logical grouping of code units (e.g., classes, modules) within a larger architectural unit.
        \item \textbf{Performance Metrics}: System-wide measurements (e.g., throughput, latency).
    \end{itemize}
 \item \textbf{Process Level}: Examines activities and operations representing workflows within the system.
    \begin{itemize}[noitemsep,topsep=0pt,leftmargin=1.5em]
        \item \textbf{Operation}: Specific tasks or actions performed within the system.
        \item \textbf{Activity}: Groups of operations into higher-level workflows.
    \end{itemize}
    
    \item \textbf{Code Level}: Focuses on the smallest units of the software system, such as classes and methods.
    \begin{itemize}[noitemsep,topsep=0pt,leftmargin=1.5em]
        \item \textbf{Class}: Foundational building blocks of object-oriented programming.
        \item \textbf{Method}: Specific functionalities or operations within a class.
    \end{itemize}

    \item \textbf{Data Level}: Focuses on stored information supporting data modeling and analysis.
    \begin{itemize}[noitemsep,topsep=0pt,leftmargin=1.5em]
        \item \textbf{Table}: Structured data collections in databases.
        \item \textbf{Entity}: Logical representations of real-world objects or concepts.
    \end{itemize}

\end{itemize}

\subsection*{Data Sources}
\begin{itemize}[noitemsep,topsep=0pt,leftmargin=*]
    \item \textbf{Open Source}: Publicly available datasets derived from open projects, repositories, or collaborative platforms.
    \item \textbf{Real World}: Data collected directly from operational systems or real-life environments.
    \item \textbf{Synthetic}: Data generated to simulate real-world scenarios when real-world data is scarce or sensitive.
\end{itemize}

\subsection*{Preprocessing Tasks}
\begin{itemize}[noitemsep,topsep=0pt,leftmargin=*]
    \item \textbf{Data Extraction}: Extracts relevant information from raw data sources (e.g., logs, traces).
    \begin{itemize}[noitemsep,topsep=0pt,leftmargin=1.5em]
        \item \textbf{Extraction of Operation and Dependency Information}: Captures workflows and dependencies.
        \item \textbf{Parsing Logs and Traces}: Extracts structured information from logs and traces.
        \item \textbf{Feature Extraction}: Identifies meaningful features for analysis.
        \item \textbf{Data Collection}: Aggregates data from diverse sources.
    \end{itemize}

    \item \textbf{Data Cleaning}: Improves data quality by removing noise and inconsistencies.
    \begin{itemize}[noitemsep,topsep=0pt,leftmargin=1.5em]
        \item \textbf{Noise Reduction}: Eliminates irrelevant or redundant information.
        \item \textbf{Normalization}: Scales and transforms data for uniformity.
    \end{itemize}

    \item \textbf{Code Analysis}: Analyzes the structural and semantic aspects of the codebase.
    \begin{itemize}[noitemsep,topsep=0pt,leftmargin=1.5em]
        \item \textbf{Semantic Analysis}: Examines the meaning and functionality of code components.
        \item \textbf{Structural Analysis}: Examines relationships and hierarchies within the codebase.
    \end{itemize}

    \item \textbf{Dependency Modeling}: Structures and transforms data to model dependencies between components.
    \begin{itemize}[noitemsep,topsep=0pt,leftmargin=1.5em]
        \item \textbf{Vectorization and Embedding}: Converts textual or categorical data into numerical formats.
        \item \textbf{Dependency Graph Generation}: Creates graphical representations of component dependencies.
        \item \textbf{Modeling and Task Structuring}: Structures data for specific migration tasks.
        \item \textbf{Matrix Structuring}: Converts dependencies into matrix formats for computational analysis.
        \item \textbf{Graph Transformation}: Transforms dependency graphs to highlight relevant relationships.
    \end{itemize}
\end{itemize}

\section*{Definitions of Key Terms for RQ3}
 \label{app:definitions_rq3}
\subsection*{Machine Learning Models}
\subsubsection{Classical Machine Learning}
\begin{itemize}
    \item \textbf{Support Vector Machines (SVM)}: Used to classify components based on their attributes, facilitating precise service decomposition.
    \item \textbf{Random Forest}: An ensemble method effective in identifying service boundaries and dependency classifications by aggregating interaction patterns.
    \item \textbf{Naive Bayes}: Useful for analysing textual data, such as code comments and documentation, to extract insights about service functionalities and dependencies.
    \item \textbf{k-Nearest Neighbors (kNN)}: Classifies components based on proximity in the feature space, aiding in optimization by grouping related entities.
    \item \textbf{Logistic Regression}: Estimates the probability of a component being part of a microservice based on extracted features.
    \item \textbf{Decision Trees}: Provide interpretable paths for classifying components based on specified rules, used for functionality and interaction-based classification.
    \item \textbf{Multilayer Perceptron (MLP)}: A simple neural network model applied to learn complex patterns for classification tasks.
\end{itemize}

\subsubsection{Deep Learning}
\begin{itemize}
    \item \textbf{Autoencoders}: Neural networks that learn to compress input data into a lower-dimensional space (encoder) and then reconstruct the original data (decoder).
    \item \textbf{Variational Autoencoders (VAE)}: Extend autoencoders by modelling the latent space probabilistically, allowing for the generation of new data similar to the training set.
    \item \textbf{Convolutional Autoencoders (CAE)}: Use convolutional layers to process images, effectively capturing spatial features.
    \item \textbf{Recurrent Networks}: Process sequential data by maintaining memory of previous inputs.
    \item \textbf{Long Short-Term Memory (LSTM)}: Addresses the vanishing gradient problem in RNNs, allowing them to capture long-range dependencies in sequences.
    \item \textbf{Gated Recurrent Units (GRU)}: Simplify LSTMs by combining forget and input gates, making them efficient for sequential tasks.
    \item \textbf{Transformers}: Use self-attention mechanisms to process sequences, enabling parallelisation and improved training speed.
    \item \textbf{BERT}: Learns word context by considering both left and right contexts, excelling in tasks like sentiment analysis.
    \item \textbf{Sentence-BERT (SBERT)}: A variation of BERT optimized for sentence-level tasks.
    \item \textbf{CodeBERT}: A transformer model designed for leveraging similarities in programming languages.
    \item \textbf{ChatGPT}: Based on the GPT architecture, supports developers in understanding and refactoring legacy systems.
\end{itemize}

\subsubsection{Graph-Based Models}
\begin{itemize}
    \item \textbf{Graph Convolutional Networks (GCN)}: Extend CNNs to graph data by aggregating features from neighbouring nodes.
    \item \textbf{Graph Attention Networks (GAT)}: Incorporate attention mechanisms into graph learning, improving representation learning.
    \item \textbf{Graph Isomorphism Networks (GIN)}: Capture graph structural properties for effective representation learning.
    \item \textbf{Relational GCN (RGCN)}: Extends GCNs to multi-relational graphs, handling various types of relationships among entities.
    \item \textbf{Variational Graph Autoencoders (VGAE)}: Combine GCNs with variational inference to learn latent representations of graphs.
\end{itemize}

\subsubsection{Reinforcement Learning}
\begin{itemize}
    \item \textbf{Fuzzy Q-Learning (FQL)}: Combines traditional Q-learning with fuzzy logic principles to handle uncertainty.
    \item \textbf{Deep Q-Learning (DQL)}: Combines traditional Q-learning with deep neural networks for handling complex environments.
    \item \textbf{Deep Deterministic Policy Gradient (DDPG)}: A model-free, off-policy reinforcement learning algorithm for continuous action spaces.
    \item \textbf{Multi-Agent Deep Deterministic Policy Gradient (MADDPG)}: Extends DDPG for multi-agent environments.
\end{itemize}

\subsection*{Learning Paradigms}
\begin{itemize}
    \item \textbf{Supervised Learning}: Trains models on labeled datasets, primarily for tasks such as anomaly detection and resource optimization.
    \item \textbf{Unsupervised Learning}: Identifies patterns and clusters in datasets without explicit labels, widely applied in microservices identification.
    \item \textbf{Semi-Supervised Learning}: Combines a small amount of labeled data with a larger pool of unlabeled data to enhance learning accuracy.
    \item \textbf{Self-Supervised Learning}: Generates pseudo-labels from raw data, allowing models to learn meaningful representations without requiring explicit labels.
    \item \textbf{Reinforcement Learning}: Trains agents to make sequential decisions by interacting with an environment and optimizing cumulative rewards.
\end{itemize}

\subsection*{Selected Features}
\subsubsection{Structural Features}
\begin{itemize}
    \item \textbf{Class Dependencies}: Include inheritance, composition, and association relationships used to analyze modularity.
    \item \textbf{Method Calls}: Reveal functional overlaps and interaction patterns between components.
    \item \textbf{Data Dependencies}: Show how data flows within a system and between its components.
    \item \textbf{Transactional Dependencies}: Identify sequences of operations that must occur together.
    \item \textbf{Call Graph Dependencies}: Capture invocation relationships between methods across components.
    \item \textbf{Graph-Based Analysis}: Uses graph structures to \revised{analyse} dependencies in microservice architectures.
    \item \textbf{Control Dependencies}: Highlight the logical flow of control within a system.
\end{itemize}

\subsubsection{Behavioral Features}
\begin{itemize}
    \item \textbf{Invocation Paths}: Represent the sequence of method calls during program execution.
    \item \textbf{Transactional Similarities}: Refer to repeated patterns of operations or interactions between components.
    \item \textbf{Response Times}: Measure the duration taken by a system to respond to a request.
    \item \textbf{Log Events}: Capture system activities for monitoring, debugging, and anomaly detection.
    \item \textbf{Contextual Log Entries}: Enhance log data with additional information.
    \item \textbf{Event Frequency}: Tracks the number of times specific events occur.
    \item \textbf{Timestamps}: Record the exact time of events.
    \item \textbf{Temporal Patterns}: Identify time-based trends in system behavior.
    \item \textbf{Fault Patterns}: Describe recurring errors or failures within the system.
\end{itemize}

\subsubsection{Performance Features}
\begin{itemize}
    \item \textbf{CPU Usage}: Tracks the percentage of processor time consumed by the system.
    \item \textbf{Memory Usage}: Measures the amount of RAM utilized.
    \item \textbf{Network Traffic}: Captures the volume of data exchanged between services.
    \item \textbf{Response Time}: Measures how quickly a system or service responds to requests.
    \item \textbf{Availability}: Measures the uptime or accessibility of a system or service.
    \item \textbf{Reliability}: Reflects a system's ability to perform without failure.
    \item \textbf{Compliance}: Ensures that the system adheres to predefined performance standards.
\end{itemize}

\subsubsection{Semantic Features}
\begin{itemize}
    \item \textbf{Semantic Embeddings}: Represent components as numerical vectors based on their contextual meaning.
    \item \textbf{Function Names}: Provide a high-level description of a method’s purpose.
    \item \textbf{Method Embeddings}: Encode methods as vectors, capturing their functionality and relationships.
    \item \textbf{API Descriptions}: Provide semantic information about system components.
    \item \textbf{Business Logic}: Captures the operational rules and processes that define system functionality.
\end{itemize}
\section*{Definitions of Key Terms for RQ4}
\label{app:definitions_rq4}
\subsection*{Classification and Prediction Metrics}
\begin{itemize}
\item \textbf{Precision:} Measures the accuracy of positive predictions.
\item \textbf{Recall:} Assesses the model's ability to identify all relevant instances.
\item \textbf{F1-Score:} Combines precision and recall for a balanced measure.
\item \textbf{Accuracy:} Represents the overall correctness of predictions.
\item \textbf{AUC:} Evaluates the quality of binary classification models.
\item \textbf{MCC:} Measures the quality of binary classifications, considering true and false positives/negatives.
\end{itemize}

\subsection*{Clustering Metrics}
\begin{itemize}
\item \textbf{Dunn Index:} Measures the separation between clusters.
\item \textbf{Silhouette Score:} Evaluates the similarity of objects within their cluster.
\item \textbf{Newman-Girvan Modularity:} Assesses the strength of network division into clusters.
\item \textbf{Non-Extreme Distribution:} Ensures balanced cluster sizes.
\item \textbf{Maximum Cluster Size:} Limits the largest cluster size to avoid granularity issues.
\item \textbf{Number of Singleton Clusters:} Counts clusters with a single element.
\end{itemize}

\subsection*{System Behavior Metrics}
\begin{itemize}
\item \textbf{Response Time:} The time taken for the system to respond to a request.
\item \textbf{Resource Utilization:} Measures the effectiveness of resource use.
\item \textbf{Energy Consumption:} Assesses the total energy required for operations.
\item \textbf{SLA Violation Rate:} The percentage of time a system fails to meet SLA obligations.
\item \textbf{Scalability:} The system's capacity to maintain or improve performance under increased workload.
\end{itemize}

\subsection*{Software Design Metrics}
\begin{itemize}
\item \textbf{Cohesion:} Evaluates the relatedness of functionalities within a service.
\item \textbf{Coupling:} Measures dependencies between services.
\item \textbf{Granularity Metric:} Evaluates service size and scope.
\item \textbf{Structural Modularity:} Measures the degree of system decomposition into independent components.
\item \textbf{Cognitive Complexity:} Assesses code understandability for developers.
\end{itemize}

\subsection*{Developer-Centric Metrics}
\begin{itemize}
\item \textbf{Developer Validation:} Assesses developer feedback on migrated services.
\item \textbf{Closeness to Manual Expert Analysis:} Compares automated classifications to expert evaluations.
\end{itemize}